\documentclass[prl,twocolumn,superscriptaddress]{revtex4-1}
\usepackage{graphicx}
\usepackage{amsmath}
\usepackage{amssymb}
\usepackage{xcolor}
\usepackage{array}

\begin{document}

\title{Direct comparison of many-body methods for realistic electronic Hamiltonians}
\collaboration{Simons collaboration on the many-electron problem} 
\noaffiliation
\author{Kiel T. Williams}
\affiliation{Department of Physics, University of Illinois at Urbana-Champaign}
\author{Yuan Yao}
\affiliation{Laboratory of Atomic and Solid State Physics, Cornell University, Ithaca, NY 14853}
\author{Jia Li}
\affiliation{Department of Physics, University of Michigan, Ann Arbor, MI 48109}
\author{Li Chen}
\affiliation{Department of Physics, University of Illinois at Urbana-Champaign}
\author{Hao Shi}
\affiliation{Center for Computational Quantum Physics, Flatiron Institute, New York, NY 10010}
\affiliation{Department of Physics, College of William and Mary, Williamsburg, VA 23185}
\author{Mario Motta}
\affiliation{IBM Almaden Research Center, 650 Harry Road, San Jose, CA 95120, USA}
\author{Chunyao Niu}
\affiliation{Department of Physics, College of William and Mary, Williamsburg, VA 23185}
\affiliation{School of Physics and Engineering, Zhengzhou University, Zhengzhou 450001, China}
\author{Ushnish Ray}
\affiliation{California Institute of Technology, Pasadena, CA 91125}
\author{Sheng Guo}
\affiliation{California Institute of Technology, Pasadena, CA 91125}
\author{Robert J. Anderson}
\affiliation{Department of Physics, King's College London, Strand, London, WC2R 2LS, U.K}
\author{Junhao Li}
\affiliation{Laboratory of Atomic and Solid State Physics, Cornell University, Ithaca, NY 14853}
\author{Lan Nguyen Tran}
\affiliation{Department of Physics, University of Michigan, Ann Arbor, MI 48109}
\affiliation{Department of Chemistry, University of Michigan, Ann Arbor, MI 48104}
\author{Chia-Nan Yeh}
\affiliation{Department of Physics, University of Michigan, Ann Arbor, MI 48109}
\author{Bastien Mussard}
\affiliation{Department of Chemistry, University of Colorado, Boulder}
\author{Sandeep Sharma}
\affiliation{Department of Chemistry, University of Colorado, Boulder}
\author{Fabien Bruneval}
\affiliation{DEN, Service de Recherches de M\'etallurgie Physique, CEA, Universit\'e Paris-Saclay, F-91191 Gif-sur-Yvette, France}
\author{Mark van Schilfgaarde}
\affiliation{Department of Physics, King's College London, Strand, London, WC2R 2LS, U.K}
\author{George H. Booth}
\affiliation{Department of Physics, King's College London, Strand, London, WC2R 2LS, U.K}
\author{Garnet Kin-Lic Chan}
\affiliation{California Institute of Technology, Pasadena, CA 91125}
\author{Shiwei Zhang}
\affiliation{Center for Computational Quantum Physics, Flatiron Institute, New York, NY 10010}
\affiliation{Department of Physics, College of William and Mary, Williamsburg, VA 23185}
\author{Emanuel Gull}
\affiliation{Department of Physics, University of Michigan, Ann Arbor, MI 48109}
\author{Dominika Zgid}
\affiliation{Department of Physics, University of Michigan, Ann Arbor, MI 48109}
\affiliation{Department of Chemistry, University of Michigan, Ann Arbor, MI 48104}
\author{Andrew Millis}
\affiliation{Center for Computational Quantum Physics, Flatiron Institute, New York, NY 10010}
\affiliation{Department of Physics, Columbia University, New York, NY 10027}
\author{Cyrus J. Umrigar}
\affiliation{Laboratory of Atomic and Solid State Physics, Cornell University, Ithaca, NY 14853}
\author{Lucas K. Wagner}
\affiliation{Department of Physics, University of Illinois at Urbana-Champaign}

\begin{abstract}
    A large collaboration carefully benchmarks 20 first principles many-body electronic structure methods on a test set of 7 transition metal atoms, and their ions and monoxides.
    Good agreement is attained between 3 systematically converged methods, resulting in experiment-free reference values. 
    These reference values are used to assess the accuracy of modern emerging and scalable approaches to the many-electron problem. 
The most accurate methods obtain energies indistinguishable from experimental results, with the agreement mainly limited by the experimental uncertainties.
Comparison between methods enables a unique perspective on calculations of many-body systems of electrons. 
\end{abstract}
\maketitle

\section{Introduction} 
A major challenge in condensed matter physics, materials physics, and chemistry is to compute the properties of electronic systems using realistic Hamiltonians.
Efficient and accurate calculations could enable computational design of drugs\cite{young2009computational} and other materials,\cite{seh_combining_2017,curtarolo_high-throughput_2013}
and shed light on a number of physical questions, such as the origin of linear-T resistivity,\cite{bruin_similarity_2013} high temperature superconductivity,\cite{muller_discovery_1987} and many other effects that currently lack satisfying explanation. 

Many-body quantum calculations on classical computers are challenging because the dimension of the Hilbert space increases dramatically with the number of particles.
For example, in the simple case of a CuO molecule with
%an almost converged (5z) basis, % the total energy is not converged in a 5z basis!
a large (5z) basis,
the Hilbert space is of dimension $10^{44}$ for the $S_z=\frac{1}{2}$ sector.
A vector of this size cannot be represented in any computer; at the present time the Oak Ridge machine Summit has approximately 250 petabytes of storage,\cite{hines_stepping_2018} which is still approximately 17 orders of magnitude too small to store a single vector.
Modern techniques therefore use compression and other techniques to approximate the state vectors.

There are many, not always mutually exclusive, approaches to dealing with the dimensionality: truncation of the wave function space through wave function \textit{ansatzes}, one-particle Green function approaches, density functional theory, Monte Carlo methods, and embedding techniques. 
The techniques vary dramatically in their computational cost and accuracy.
Most studies\cite{bauschlicher_theoretical_1995,furche_performance_2006,doblhoff-dier_diffusion_2016,verma_assessment_2017,xu_practical_2015,minenkov_troubles_2016,thomas_accurate_2015, Shee-TMD-JCTC_15_2346_2019,shee_phaseless_2018} judge the accuracy of the methods by comparing to experimental energies,\cite{page_completing_1990} which are computed by taking differences of total energies and are therefore subject to fortuitous cancellation of error.
Instead, in this study we include 3 systematically improvable methods with sufficiently small prefactors that they yield almost exact total energies within the chosen basis set and serve as a benchmark for testing all other methods.

In this manuscript, we apply a diverse array of 20 established and emerging techniques to a test set of small, realistic transition metal molecules and atoms. 
Each technique was implemented by an expert, and employed precisely the same Hamiltonian.
This approach allows us to directly assess methodological differences without confounders such as different Hamiltonians,
and has been important for a previous benchmark study of the hydrogen chain.\cite{Motta17} 
For these systems, we achieve convergence of exponentially scaling but systematically convergeable methods at the order of 1 mHartree in the \textit{total energy}, or about 300 K, establishing a reliable reference on realistic Hamiltonians with complex atoms.
We then assess the accuracy of more approximate approaches for computing the total energy of atoms and molecules, which allows some assessment of transferability of performance with increasing system size.
Finally, we study how errors in the total energies translate into errors of physical observables obtained as differences of total energies, and we make comparisons to experiments.
These results provide an important reference for the development of techniques that can address the larger goal of computing electronic properties of realistic materials.

\section{Methodology} 

Table~\ref{table:abbreviations} lists the methods tested in this work. 
It includes most of the common techniques to address the many-electron problem, as well as some emerging methods. 
It also includes a few methods such as CISD which are no longer commonly used but have historical relevance. 
The methods in this benchmark vary dramatically in their computational cost; the density functional theory methods required only a few minutes to complete the test set, while some of the more advanced techniques were not able to treat every basis for every system with the available amount of computer time.
The methods also scale very differently, ranging from $\mathcal{O}(N_e^3)$ to exponential in the number of electrons $N_e$.
Of the 3 systematically converged methods (iFCIQMC, DMRG and SHCI) only SHCI was performed for all the systems in all the basis sets.  
Consequently, SHCI energies will be used as the reference.

Some of the other techniques are in principle systematically improvable, such as configuration interaction, coupled cluster, self-energy embedding theory, and the Monte Carlo methods, but convergence to better than 1 mHa was not achieved on these systems for the level of the method employed.
Some of the techniques give upper bounds to the exact energy, such as DMC, CISD, DMRG, and HF.
Finally, for completeness it should be noted that the methods also require different levels of specification to define the approximations used. 
For example, some of the methods can be reproduced only by specifying the initial starting determinant; others require defining an initial multideterminantal wavefunction, or the choice of partitioning between high-level and low-level methods.

\begingroup
\squeezetable
\begin{table}[!h]
\caption{A list of abbreviations used in this benchmark.
Details are available in the Supplementary Material.
Column A lists the largest basis set employed by that method for at least one of the transition metal atoms,
Column B lists the same for the monoxide molecules.
The basis sets are abbreviated in order as d, t, q, 5, and c for complete basis set.
}\label{table:abbreviations}
\begin{ruledtabular}
\begin{tabular} {l|>{\raggedright}p{4.5cm}|c|c}
Abbreviation & Method & A & B \\
\hline
AFQMC(MD) & Auxiliary field quantum Monte Carlo with a multi-determinant trial function\cite{ShiweiPhaseless,Mario-WIRES-2018} & 5 & 5\\
B3LYP & DFT with the B3LYP functional\cite{becke_densityfunctional_1993} &5 & 5\\
CISD & Configuration interaction with singles and doubles & 5 & 5 \\
DMC(SD)& Fixed node diffusion Monte Carlo with a single determinant nodal surface\cite{foulkes_quantum_2001,wagner_qwalk:_2009} & c & c\\
DMRG & Density matrix renormalization group\cite{olivares2015ab,sharma2012spin}& t & d \\
GF2 & Second order Green function\cite{Phillips14,Rusakov16} & q & q \\
HF & Hartree-Fock & 5 & 5\\
HF+RPA & Hartree-Fock random phase approximation\cite{Eshuisa10} &t & t \\
HSE06 & DFT with the HSE06 functional\cite{HSE03,heyd_erratum:_2006} &5 & 5\\
iFCIQMC & Initiator full configuration interaction quantum Monte Carlo\cite{doi:10.1063/1.3193710,doi:10.1063/1.3302277} & q & d\\
LDA & DFT in the local density approximation\cite{ceperley_ground_1980,vosko_accurate_1980} & 5 & 5\\
MRLCC & Multireference localized coupled cluster\cite{mrlcc1,mrlcc2,mrlcc3,mrlcc4} &5 & 5\\ 
PBE & DFT in the PBE\cite{perdew_generalized_1996} approximation &5 & 5  \\
QSGW & Quasiparticle self-consistent GW approximation\cite{Faleev04} & t & t \\
SCAN & DFT with SCAN functional\cite{sun_strongly_2015} & 5 & 5\\
SC-GW & Self-consistent GW approximation \cite{Hedin65,Tran17B}& t & - \\
SEET(FCI/GF2) & Self-energy embedding theory with many-body expansion.\cite{Tran15b,Tran17,Zgid17,Tran18,Rusakov18}& q & q \\
SHCI & Semistochastic heatbath configuration interaction\cite{HolTubUmr-JCTC-16,LiOttHolShaUmr-JCP-18} & 5 &5 \\
UCCSD & Unrestricted coupled cluster with singles and doubles\cite{bartlett_coupled-cluster_2007} & 5 &5 \\
UCCSD(T) & Unrestricted coupled cluster with singles, doubles, and perturbative triples\cite{bartlett_coupled-cluster_2007} & 5 & 5 \\
    \end{tabular}
\end{ruledtabular}
\end{table}
\endgroup

We consider transition metal systems, with the core electrons removed using effective core potentials\cite{trail_pseudopotentials_2013,trail_correlated_2015,trail_shape_2017}. 
These potentials accurately represent the core\cite{bennett_new_2017} in many-body simulations and allow all the methods considered in this work to use the same Hamiltonian.
In addition, they provide an easy way to include scalar relativistic effects, needed for meaningful comparison to experiment.
These potentials are available for O, Sc, Ti, V, Cr, Mn, Fe, and Cu, which defines our test set. 
We consider these atoms, their ions, and the corresponding transition metal monoxide molecules.
To simplify the comparison, the molecules were computed at their equilibrium geometry.

Almost every electronic structure method (all the methods in this study except DMC) works in a finite basis. 
Here, we follow the chemistry convention of defining an ascending basis set denoted by the z($\zeta$) value, ranging from 2 to 5; i.e., dz, tz, qz, and 5z. 
For each system, we consider the first principles Hamiltonian projected onto the basis, making for a total of $23 \times 4 = 98$ calculations for each method. 
See the Supplementary Material for details on the precise basis sets used in this study.
While the results are only comparable to experiment in the complete basis set limit (cbs), for each basis set there corresponds a projected Hamiltonian, which also has an exact solution. 
We thus can compare methods \textit{within a basis} since the Hamiltonian is defined precisely. 

In Table~\ref{table:abbreviations}, we list the methods considered in this work.
The deviation in the total energy between two methods $m$ and $n$ is computed as
\begin{equation}
    \sigma(m,n) = \sqrt{\frac{\sum_{i \in \text{systems} } (E_i(n)-E_i(m))^2}{N}},
    \label{eqn:rms}
\end{equation}
where $N$ is the total number of calculations performed in common between the methods.
This is a measure of how well the output total energies between two methods agree.  
It is possible for two methods with large $\sigma$ to agree on energy \textit{differences} if there is significant cancellation of errors. 

To compare total energies between methods and systems in a consistent way, we use the concept of percent of correlation, commonly used in quantum chemistry:
\begin{equation}
    \text{\% correlation energy}(m) = 100 \times \frac{E_{\rm HF} - E_{m}}{E_{\rm HF}-E_{\rm SHCI}},
    \label{eqn:correlation_energy}
\end{equation}
where $E_{\rm HF}$ is the Hartree-Fock energy, $m$ stands for the method under consideration, and $E_{\rm SHCI}$ is the total energy computed in the basis by the SHCI method.
At 100\% of the correlation energy, the exact result is obtained. 
This quantity is particularly useful since methods tend to obtain similar percentages of the correlation energy across different basis sets and systems.

Extrapolation to the basis set limit is done making the usual assumption that the correlation energy
(difference between Hartree-Fock and the exact energy) scales as $1/n^3$, where $n$ is the cardinal number of the basis set, and that the Hartree-Fock energy exponentially converges to the complete basis limit.
Complete basis set extrapolation is necessary for comparison of the finite basis set results to experiment, diffusion Monte Carlo (DMC), and density functional theory results.
DMC works directly in the complete basis limit, whereas density functional methods are designed to reproduce complete basis set limit energies.
The uncertainty in the extrapolation, judged from the variation between different fits to the extrapolation, is approximately 2--4 mHa; for details, see the Supplementary Material.
Thus, in this test set, the largest uncertainty in the complete basis set total energy is due to the extrapolation of finite basis set energies to the infinite limit.

The energy differences studied are the ionization potential of a transition metal atom $M$: $\text{IP}=E(M^+) - E(M)$, and the dissociation energy of a metal oxide molecule $MO$: $\text{DE}=E(M)+E(O)-E(MO)$.
These quantities have been studied in detail for these systems in the past, for example Refs~\cite{bauschlicher_theoretical_1995,furche_performance_2006,doblhoff-dier_diffusion_2016,verma_assessment_2017,xu_practical_2015,bligaard_toward_2016,mardirossian_thirty_2017,minenkov_troubles_2016,thomas_accurate_2015,tew_explicitly_2016,johnson_communication:_2017}, among others. 
However, none of these previous studies have attained reference energies as well-converged as the ones in this paper,
and none compare energies from a large number of methods.

\section{Results}

\begin{figure}
\includegraphics{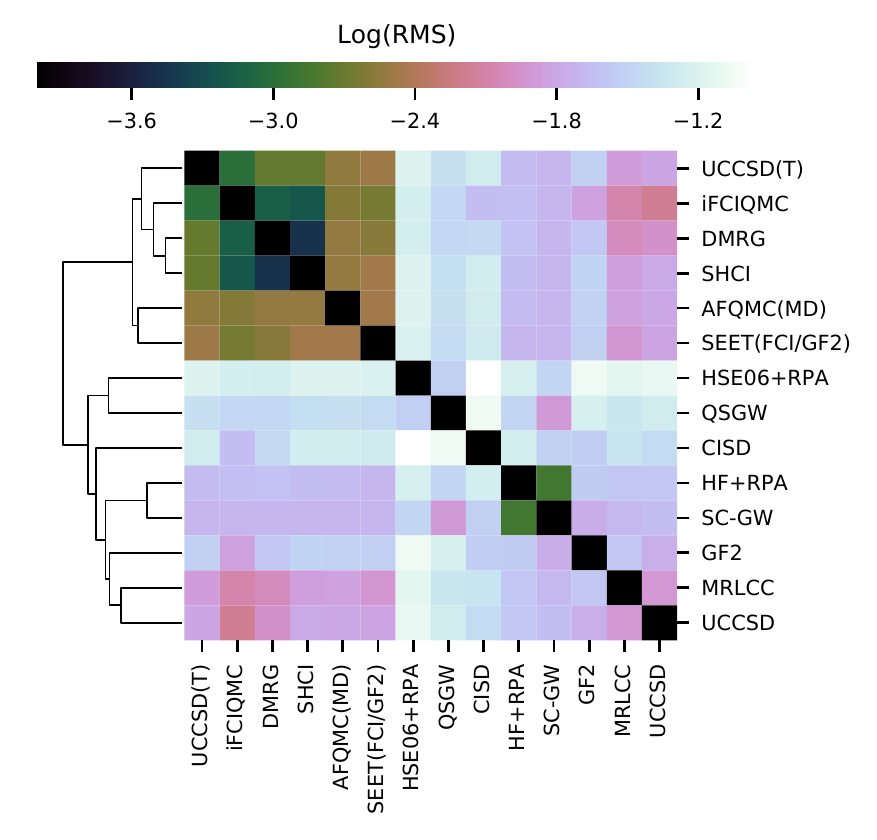}
\caption{Cluster analysis of electronic structure methods in this work. 
The matrix values are the logarithm of the RMS deviation of the total energy in Hartrees (Eq.~\ref{eqn:rms}) between the two methods.
}
    \label{fig:clustermap}
\end{figure}

\begin{figure*}[htb]
\includegraphics{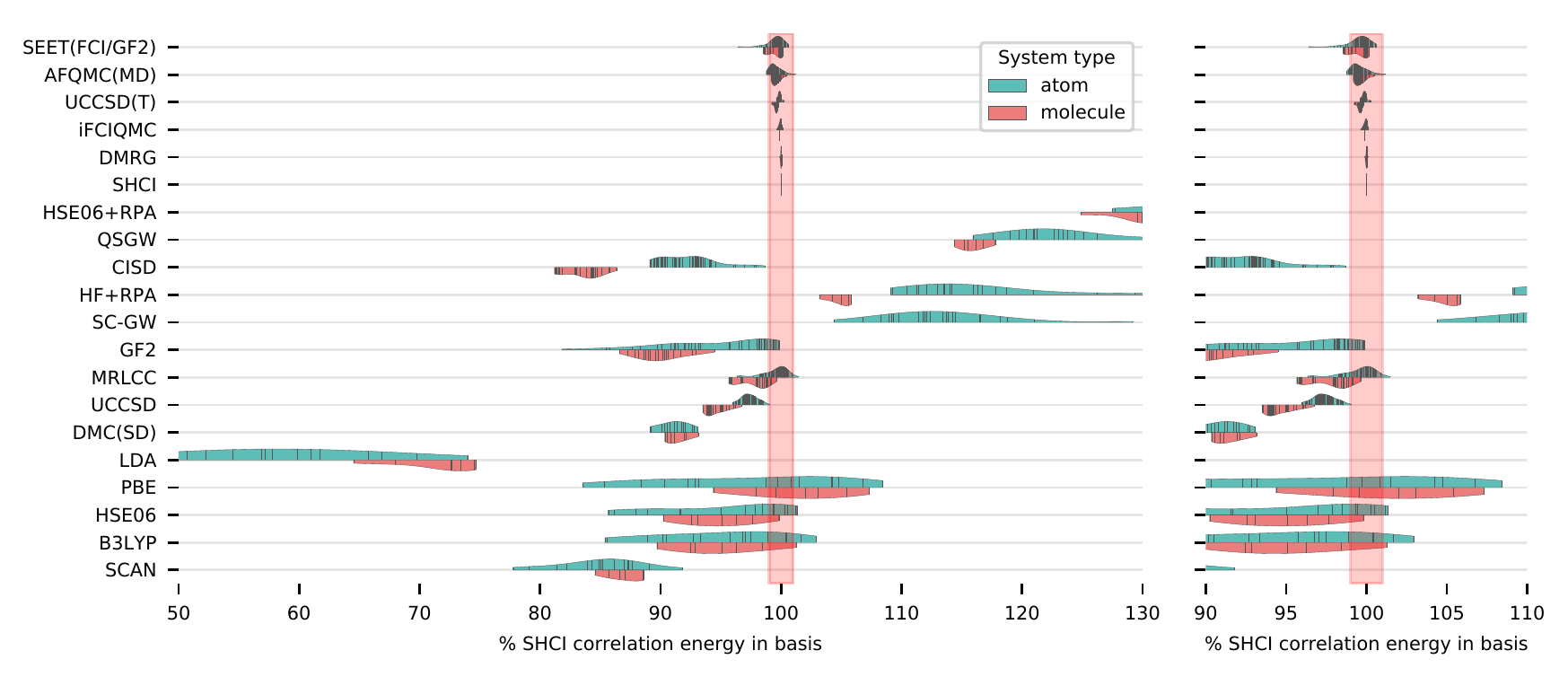}
\caption{Kernel density estimation\cite{rosenblatt_remarks_1956,parzen_estimation_1962,michael_waskom_mwaskom/seaborn:_2018} of the percent of the SHCI-computed correlation energy within each basis obtained by each of the methods in the benchmark set.
All basis sets available are plotted; individual data points are indicated by small lines.
}\label{fig:correlation_energy}
\end{figure*}

\begin{figure*}
    \includegraphics{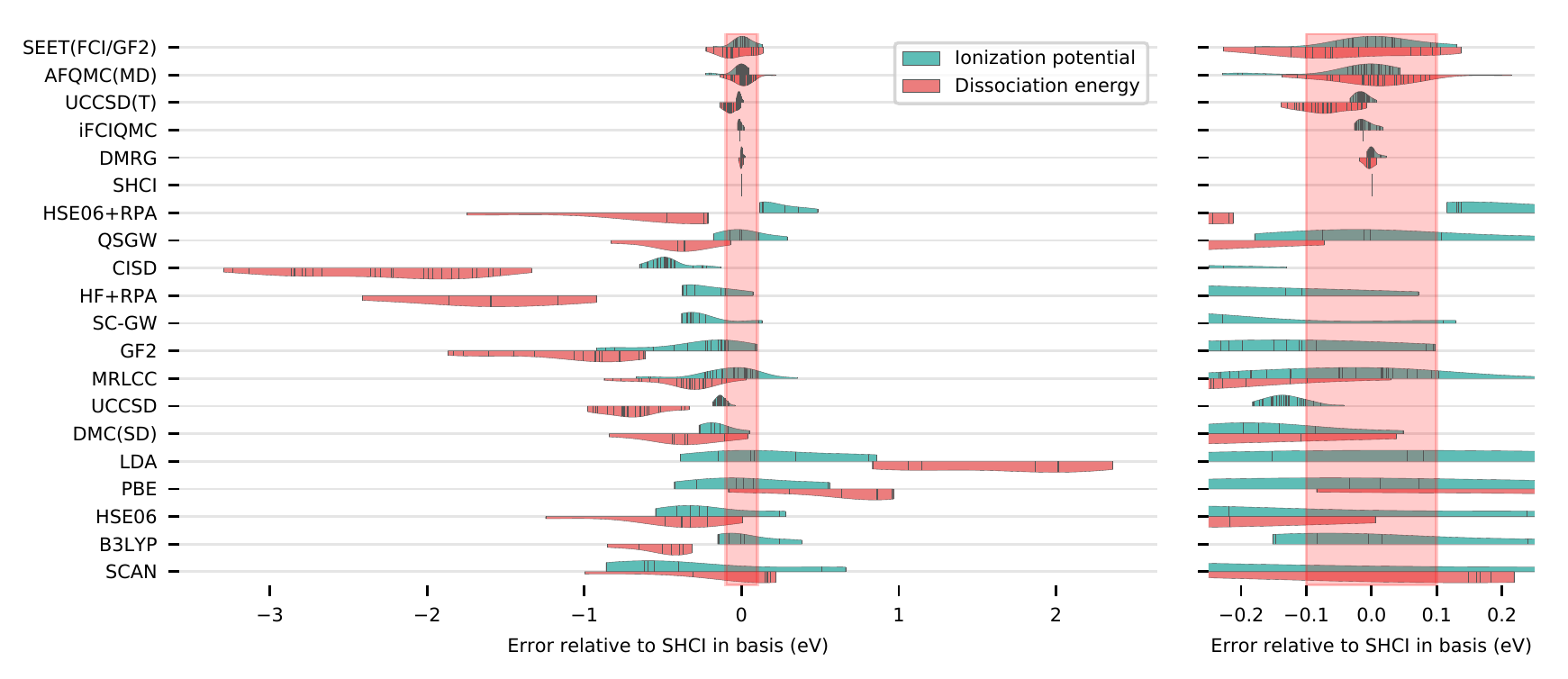}
    \caption{Kernel density estimation plot of dissociation energy and ionization potential of molecules and atoms to SHCI reference calculations. 
	Methods are ordered according to the clustering in Fig~\ref{fig:clustermap}.
 }\label{fig:be_ip}
\end{figure*}

\begin{figure}
    \includegraphics{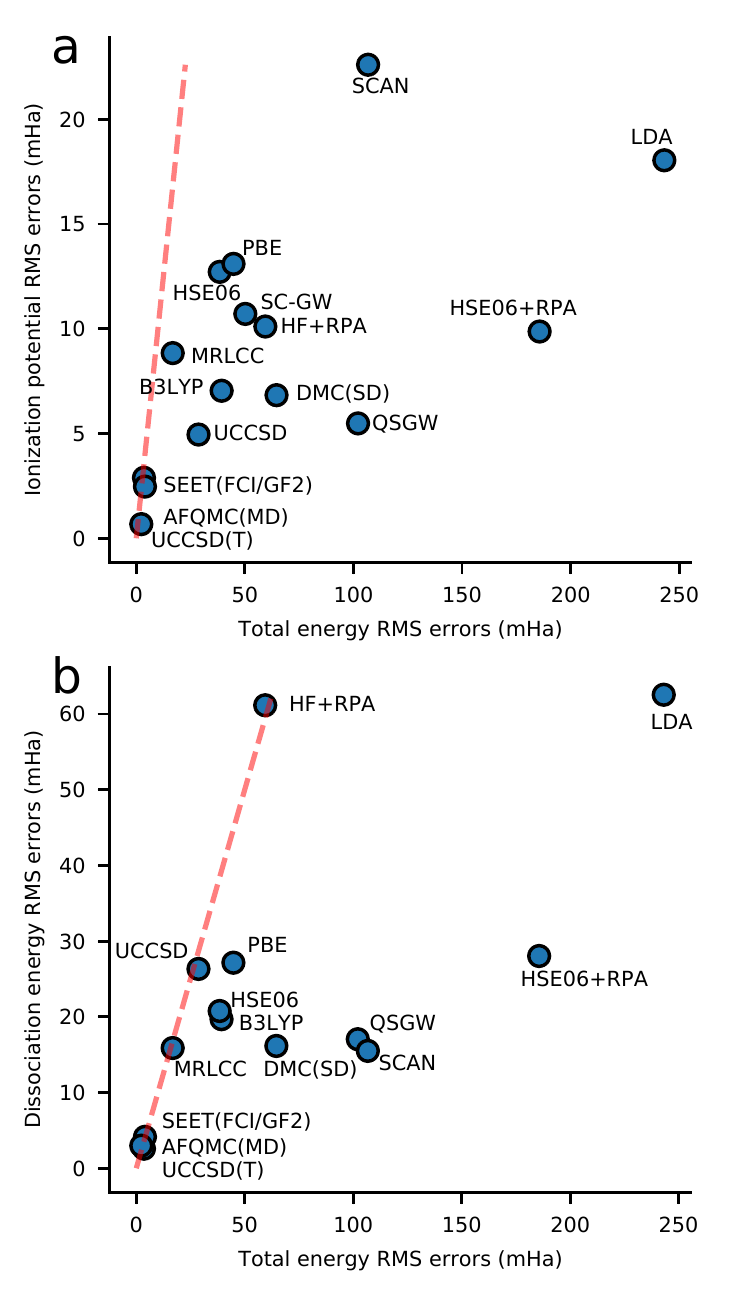}
    \caption{Cancellation of error for many methods in this study, computed by comparing the RMS error in the total energy to the RMS error in the (a) ionization energy of the atoms and (b) dissociation energy for molecules. HF and CISD were excluded from the comparison for more detail in the more accurate methods; they are off the scale here. The red dashed line corresponds to no cancellation of error. } 
    \label{fig:cancellation}
\end{figure}

\begin{figure}
    \includegraphics{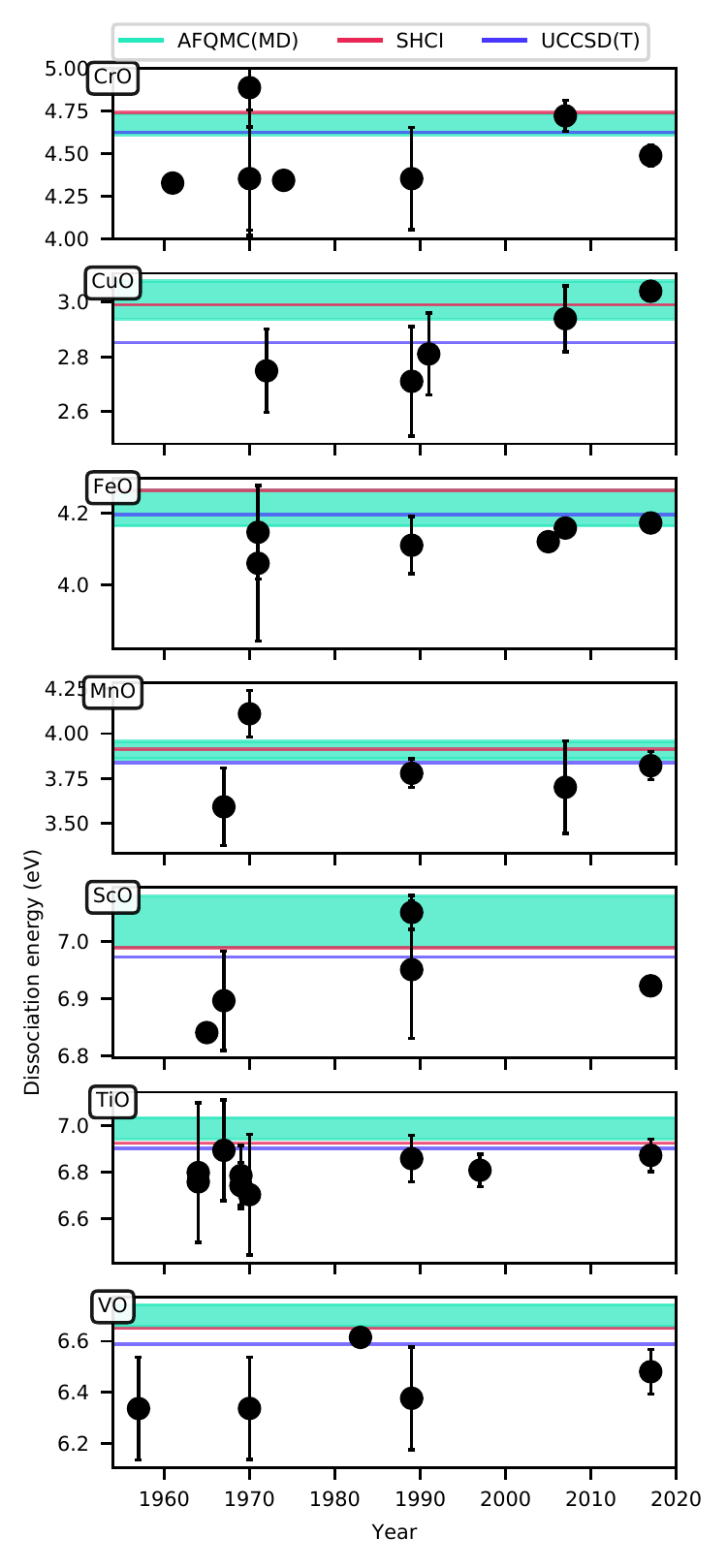}
    \caption{Comparison of 5z dissociation energies of the transition metal molecules obtained from the more accurate methods used in this work to experiment.
    The x-axis is the year the experimental result was published and the width of the bars indicate statistical/systematic uncertainties.
    }
    \label{fig:experiment}
\end{figure}

We show several views of the data collected in this study in the figures.
The Supplementary Material contains various tables and the complete set of data ($\sim$1200 calculations) on which these plots are based. 
Fig~\ref{fig:clustermap} establishes that several high accuracy techniques are in agreement and establishes a reference technique SHCI.
Fig~\ref{fig:correlation_energy} compares the performance of methods in computing the total energy as compared to the reference.
Fig~\ref{fig:be_ip} compares the performance of methods in computing ionization potential of the atoms and dissociation energy of the molecules. 
Fig~\ref{fig:cancellation} summarizes the cancellation of error for different techniques in computing the differences in energies. 
Finally, Fig~\ref{fig:experiment} compares calculations using methods found to be accurate to the experimental dissociation energies. 
In this section, we examine related methods in the context of these different views.

In Fig~\ref{fig:clustermap}, we show a cluster analysis of the total energies using Eq.~\ref{eqn:rms}, evaluated on the intersection of basis sets and systems available for both methods, as the distance metric.
iFCIQMC, DMRG and SHCI were converged to very high levels of accuracy.
In fact these three methods agree to $\sim$1 mHa for all systems and basis sets that were computed.
Because of this 3-fold agreement, we can take any of these results as the exact ground state energy in a given basis set to within an RMS error of less than 1 mHa, which is approximately what is termed ``chemical accuracy'' in the context of energy differences. 
Here we have achieved 1 mHa accuracy in the total energy of the ground state.
However, as shown in Table~\ref{table:abbreviations}, iFCIQMC and DMRG calculations were feasible within the available computer time for only the smaller basis sets, so we use SHCI as the reference.
For finite basis sets, the estimated uncertainty is $\sim$ 1 mHa, and for the complete basis set, the estimated uncertainty is $\sim$ 2-4 mHa due to the extrapolation uncertainty.

Density functional methods have a large spread across systems in the percent of correlation energy attained (Fig~\ref{fig:correlation_energy}). 
The gradient corrected and the hybrid functionals (B3LYP, HSE06, PBE and SCAN) improve the LDA.
The most recently proposed of these, SCAN, is more consistent in the percent of correlation energy obtained at around 80--90\% of the correlation energy.
Fig.~\ref{fig:cancellation} shows that it also benefits more than the other functionals from a cancellation of errors between the atom and the molecule to give more accurate dissociation energies, although it has less cancellation of errors for the ionization potentials.
Much of the improvement in accuracy of the hybrid functionals over PBE is in the cancellation of error.

The random phase approximation (RPA) and both versions of GW  overestimate the correlation energy as shown in Fig~\ref{fig:correlation_energy}.
While the total energy tends to be too low, those errors tend to cancel for QSGW applied to energy differences, as can be seen in Figs.~\ref{fig:be_ip} and \ref{fig:cancellation}. 

As can be seen in Fig~\ref{fig:correlation_energy}, configuration interaction with singles and doubles (CISD), a truncated determinant expansion technique well-known to have size consistency defects, performs much better for the atoms than the molecules, which leads to rather poor predictions for the dissociation energy of the molecules (Fig~\ref{fig:be_ip}).
The error is large enough that CISD was not included in Fig~\ref{fig:cancellation} to improve readability of the more accurate numbers.
We note that unrestricted coupled cluster with singles and doubles (UCCSD), which is size consistent, also performs worse on the molecules than the atoms, though to a lesser degree than CISD.
This results in underestimation of the dissociation energy (Fig~\ref{fig:be_ip}), and no cancellation of error in the dissociation energy, but significant cancellation in the ionization potential (Fig~\ref{fig:cancellation}).

Fixed node diffusion Monte Carlo with a single determinant trial function (DMC(SD)) yields a lower bound to the extrapolated correlation energy, corresponding to an upper bound to the total energy, which is apparent in Fig~\ref{fig:correlation_energy}. 
The remaining energy is the fixed node error, the main approximation in the DMC calculations, which for a single Slater determinant nodal surface is much larger than the extrapolation uncertainty.
With the single Slater determinant, DMC obtains 90--95\% of the correlation energy quite consistently, in line with previous benchmarks on smaller systems\cite{brown_energies_2007}. 
This consistency results in significant cancellation of error (Fig~\ref{fig:cancellation}) in the dissociation energy and ionization potential.

Self energy embedding theory with a full configuration interaction solver and GF2 embedding (SEET(FCI/GF2)) obtains results in good agreement with the reference total  energy (Fig~\ref{fig:correlation_energy}), resulting in accurate energy differences (Fig~\ref{fig:be_ip}).
Consequently, it lies very close to the $x=y$ line in Fig~\ref{fig:cancellation} and does not benefit from additional cancellation of error, as the energies are already accurate. 
The errors in the total energy are not strongly correlated with the atomic species; for example the error in the Ti atom is not
statistically similar to the error in the TiO atom, resulting in little cancellation of error.

The auxilliary field quantum Monte Carlo with a multiple determinant trial function (AFQMC(MD)) gives good agreement with the reference total  energy, with an RMS deviations of about 3 to 4 mHa. 
The dissociation energies have an RMS deviation of $\sim 2.5$mHa, which is consistent with the conclusion of a recent benchmark on a large set of transition metal diatomics \cite{Shee-TMD-JCTC_15_2346_2019}. 
The use of single determinant UHF trial wave functions would lead to less accurate results, roughly doubling the RMS error in the total energy of the molecules (see Supplementary Material I.A).

Coupled cluster with singles, doubles, and perturbative triples (UCCSD(T)) performs very well on these systems, obtaining close to 100\% of the correlation energy.
For these problems, UCCSD(T) has a notably low cost for high performance.
The accuracy of UCCSD(T) is likely due to the fact that these systems are not strongly multi-reference, in that even in the near-exact wave functions, there is a single dominant determinant that makes a large contribution to the wave function.
This can be seen by examining the natural orbital occupations; for example in UCCSD, the spin-resolved natural orbitals with large occupations have occupations of 0.96 or greater. 
The single reference nature also explains the mediocre performance of the multi-reference methods such as MRLCC, which sacrifice some accuracy in the single reference case to treat multi-reference situations more accurately.
In general, active space techniques, which operate within an explicitly chosen subspace of the larger Hilbert space, are not very effective for these systems.

We believe that the reference data produced computationally has lower uncertainties than experiment for the purposes of benchmarking quantum calculation techniques.
The ionization potential of the large-basis SHCI results is in agreement with experiment with mean absolute deviation of 0.2 mHa, or 7 meV, so one could equivalently use experiment or the SHCI reference values, as can be verified in Table VI of the Supplementary Material. 
The experimental dissociation energy estimation is limited by the challenges of the measurements and the experimental measures differ from one another by as much as 0.5 eV.
In Fig~\ref{fig:experiment}, the high accuracy  estimates of the dissociation energy of the molecules is shown, compared to experimental values with zero point energy removed\cite{merer_spectroscopy_1989,carlson_electronic_1965,darwent_bond_1970,carlson_wavefunctions_1964,berkowitz_thermodynamics_1957,balducci_dissociation_1971,clemmer_reaction_1991}.
For these systems, the experimental uncertainty of the dissociation energy is larger than the difference between the most accurate techniques in this benchmark.
Remarkably, SHCI, UCCSD(T), and AFQMC(MD) agree to about 0.1 eV for all the molecules.
We also should note that since we used effective core potentials to standardize the benchmark, there may be some small errors in comparing directly to experiment.
However, we see no evidence that the potentials used are limiting the accuracy; the most accurate methods obtain results well within the experimental uncertainty, with the possible exception of VO, for which most of the experimental values are slightly below the theoretical ones.

When computing differences of total energies, both methodological errors and errors due to finite basis sets tend to cancel.
In Fig~\ref{fig:cancellation} we quantify the methodological cancellation of errors in many of the techniques studied in this work.
Considering basis set errors, the RMS error in the total energy in the commonly used tz basis compared to the complete basis set limit is 75 mHa, while the RMS error in the ionization energy and dissociation energy for the same comparison are 1.6 mHa and 6 mHa respectively, as can be seen in Table VIII in the Supplementary information.

\section{Conclusion}

We surveyed 20 advanced many-electron techniques on precisely defined realistic Hamiltonians for transition metal systems. 
For a given basis set, we achieved $\sim$ 1 mHa agreement on the total energy between high accuracy methods, which provides a \textit{total energy} benchmark for many-body methods. 
To our knowledge, such an agreement is unprecedented for first principles calculations of transition metal systems.
Our accurate reference energies should enable the development of approximate, but more computationally efficient, many-body techniques as well as better density functionals, without the necessity of experimental reference values.
These systems are also a useful test for future quantum computing algorithms.
To enable such comparisons, we include \texttt{pyscf} scripts that can execute the benchmark for any density functional available in libxc\cite{marques_libxc:_2012}, and can export the one- and two-body integrals needed for testing many-body methods.

We have assessed the state of the art in achieving high accuracy in realistic systems. 
The benchmark set includes systems with large Hilbert spaces of around 10$^{44}$ determinants.
While these spaces are so large that a single vector cannot fit in any computer memory, the computations are feasible due to powerful compression of that space.
The systematically converged techniques used in this work (DMRG, FCIQMC, and SHCI), were able to achieve excellent agreement, but can be applied only to relatively small systems due to their computational cost.
It is thus important to understand the errors in lower-scaling techniques that can be applied to larger systems, and whether performance on small systems is transferable to larger systems. 
Our study takes a step in that direction, since we were able to achieve converged results for both correlated atoms and molecules, and indeed we observed that the accuracy of some techniques degrades with system size.

To avoid misinterpretation of the results, we make a comment here.
In order to ensure high quality results, it was necessary to limit the number of systems on which this benchmark was performed. 
While treating electron correlation accurately is important to obtain accurate results, these systems have a particular character of correlation.
In a determinant expansion of the wave function, the systems chosen here have one determinant with a large weight and many determinants with small weights, rather than several determinants with large weights. 
For such systems, methods such as UCCSD(T) are accurate.
The performance profile will likely be different for differently correlated chemical systems, so benchmarking efforts of similar quality in that realm would be highly valuable.

This work was supported by a grant from the Simons Foundation as part of the Simons Collaboration on the Many Electron Problem.

\bibliography{realistic}
\end{document}

\title{Supplementary Material: Direct comparison of many-body methods for realistic electronic Hamiltonians}

\begin{sidewaystable}
\caption{Calculations in the database. In each cell, the symbols correspond to the basis performed as follows: 
d: vdz, t: vtz, q: vqz, 5: v5z, c: cbs.  A dash means that there were no calculations for that system using that technique. } 
\begin{ruledtabular}
\begin{tabular}{llllllllllllllllllllllll}
{} & \multicolumn{23}{c}{System} \\
Method &    Cr &   Cr+ &   CrO &    Cu &   Cu+ &   CuO &    Fe &   Fe+ &   FeO &    Mn &   Mn+ &   MnO &     O &    O+ &    Sc &   Sc+ &   ScO &    Ti &   Ti+ &   TiO &     V &    V+ &    VO \\
\hline
AFQMC(MD)     &  dtq5 &  dtq5 &  dtq5 &  dtq5 &   dtq5 &  dtq5 &  dtq5 &  dtq5 &  dtq5 &  dtq5 &  dtq5 &  dtq5 &  dtq5 &  dtq5 &  dtq5 &  dtq5 &  dtq5 &  dtq5 &  dtq5 &  dtq5 &  dtq5 &  dtq5 &  dtq5 \\
B3LYP         &  dtq5 &  dtq5 &  dtq5 &  dtq5 &  dtq5 &  dtq5 &  dtq5 &  dtq5 &  dtq5 &  dtq5 &  dtq5 &  dtq5 &  dtq5 &  dtq5 &  dtq5 &  dtq5 &  dtq5 &  dtq5 &  dtq5 &  dtq5 &  dtq5 &  dtq5 &  dtq5 \\
CISD          &  dtq5 &  dtq5 &  dtq5 &  dtq5 &  dtq5 &  dtq5 &  dtq5 &  dtq5 &  dtq5 &  dtq5 &  dtq5 &  dtq5 &  dtq5 &  dtq5 &  dtq5 &  dtq5 &  dtq5 &  dtq5 &  dtq5 &  dtq5 &  dtq5 &  dtq5 &  dtq5 \\
DMC(SD)       &     c &     c &     c &     c &     c &     c &     c &     c &     c &     c &     c &     c &     c &     c &     c &     c &     c &     c &     c &     c &     c &     c &     c \\
DMRG          &    dt &    dt &     d &    dt &    dt &     d &    dt &    dt &     d &    dt &    dt &     d &    dt &    dt &    dt &    dt &     d &    dt &    dt &     d &    dt &    dt &     d \\
GF2           &   dtq &   dtq &   dtq &   --- &   --- &   dtq &   dtq &   dtq &   dtq &   dtq &   dtq &    dt &   dtq &   dtq &   dtq &   dtq &   dtq &   dtq &   dtq &   dtq &   dtq &   dtq &   dtq \\
HF+RPA        &     t &     t &     t &     t &     t &   --- &    dt &     t &     t &     t &     t &     t &   dtq &   dtq &     t &     t &     t &     t &     t &     t &     t &     t &     t \\
HSE06         &     dtq5 &     dtq5 &     dtq5 &     dtq5 &     dtq5 &   --- &    ddtq5 &     dtq5 &     dtq5 &     dtq5 &     dtq5 &     dtq5 &   ddtq5q &   ddtq5q &     dtq5 &     dtq5 &     dtq5 &     dtq5 &     dtq5 &     dtq5 &     dtq5 &     dtq5 &     dtq5 \\
LDA           &  dtq5 &  dtq5 &  dtq5 &  dtq5 &  dtq5 &  dtq5 &  dtq5 &  dtq5 &  dtq5 &  dtq5 &  dtq5 &  dtq5 &  dtq5 &  dtq5 &  dtq5 &  dtq5 &  dtq5 &  dtq5 &  dtq5 &  dtq5 &  dtq5 &  dtq5 &  dtq5 \\
MRLCC         &  dtq5 &  dtq5 &  dtq5 &  dtq5 &  dtq5 &  dtq5 &  dtq5 &  dtq5 &  dtq5 &  dtq5 &  dtq5 &  dtq5 &  dtq5 &  dtq5 &  dtq5 &  dtq5 &  dtq5 &  dtq5 &  dtq5 &  dtq5 &  dtq5 &  dtq5 &  dtq5 \\
PBE           &  dtq5 &  dtq5 &  dtq5 &  dtq5 &  dtq5 &  dtq5 &  dtq5 &  dtq5 &  dtq5 &  dtq5 &  dtq5 &  dtq5 &  dtq5 &  dtq5 &  dtq5 &  dtq5 &  dtq5 &  dtq5 &  dtq5 &  dtq5 &  dtq5 &  dtq5 &  dtq5 \\
PBE+RPA       &     t &     t &     t &     t &     t &   --- &     t &     t &     t &     t &     t &     t &   dtq &   dtq &     t &     t &     t &     t &     t &     t &     t &     t &     t \\
QSGW          &     t &     t &     t &   --- &   --- &   --- &    dt &     t &     t &     t &     t &     t &   dtq &   dtq &     t &     t &     t &     t &     t &     t &     t &     t &     t \\
SC-GW         &     d &     d &   --- &   --- &   --- &   --- &     d &     d &   --- &     d &     d &   --- &     d &     d &     d &     d &   --- &     d &     d &   --- &     d &     d &   --- \\
SCAN          &  dtq5 &  dtq5 &  dtq5 &  dtq5 &  dtq5 &  dtq5 &  dtq5 &  dtq5 &  dtq5 &  dtq5 &  dtq5 &  dtq5 &  dtq5 &  dtq5 &  dtq5 &  dtq5 &  dtq5 &  dtq5 &  dtq5 &  dtq5 &  dtq5 &  dtq5 &  dtq5 \\
SEET(FCI/GF2) &   dtq &   dtq &    dq &   --- &   --- &   dtq &   dtq &   dtq &   dtq &   dtq &   dtq &   --- &   dtq &   dtq &   dtq &   dtq &   dtq &   dtq &   dtq &   dtq &   dtq &   dtq &   dtq \\
SHCI          &  dtq5 &  dtq5 &  dtq5 &  dtq5 &  dtq5 &  dtq5 &  dtq5 &  dtq5 &  dtq5 &  dtq5 &  dtq5 &  dtq5 &  dtq5 &  dtq5 &  dtq5 &  dtq5 &  dtq5 &  dtq5 &  dtq5 &  dtq5 &  dtq5 &  dtq5 &  dtq5 \\
UCCSD         &  dtq5 &  dtq5 &  dtq5 &  dtq5 &  dtq5 &  dtq5 &  dtq5 &  dtq5 &  dtq5 &  dtq5 &  dtq5 &  dtq5 &  dtq5 &  dtq5 &  dtq5 &  dtq5 &  dtq5 &  dtq5 &  dtq5 &  dtq5 &  dtq5 &  dtq5 &  dtq5 \\
UCCSD(T)      &  dtq5 &  dtq5 &  dtq5 &  dtq5 &  dtq5 &  dtq5 &  dtq5 &  dtq5 &  dtq5 &  dtq5 &  dtq5 &  dtq5 &  dtq5 &  dtq5 &  dtq5 &  dtq5 &  dtq5 &  dtq5 &  dtq5 &  dtq5 &  dtq5 &  dtq5 &  dtq5 \\
iFCIQMC       &   dtq &   dtq &   --- &   dtq &   dtq &   --- &   dtq &   dtq &   --- &   dtq &   dtq &   --- &   dtq &   dtq &   dtq &   dtq &     d &   dtq &   dtq &   --- &   dtq &   dtq &   --- \\
\end{tabular}
\end{ruledtabular}
\end{sidewaystable}

\section{Detailed methodology} 

\label{sec:methods}

\subsection{AFQMC}

The AFQMC method  \cite{ShiweiConstraint, ShiweiPhaseless, AFQMC-lecture-notes-2013} estimates the ground-state properties of a many-fermion system by statistically sampling the wave function $|\psi_g\rangle \propto e^{-\beta \hat{H}} |\psi^0\rangle$, where $|\psi^0\rangle$  is an initial wave function which is nonorthogonal to the ground state.
The projection is carried out iteratively by small time step $ e^{-\Delta \tau \hat{H}}$, with $\beta = n \Delta \tau$
sufficiently large to project out all excited states.
The propagator is represented as $e^{-\Delta \tau \hat{H}} = \int d \boldsymbol{x} p(\boldsymbol{x} ) \hat{B}(\boldsymbol{x} )$
where $\hat{B}(\boldsymbol{x} )$ is a one-body operator which depends on
the vector $\boldsymbol{x} $, and $p(\boldsymbol{x} )$ is a probability distribution.  This representation maps a many-body system into  an ensemble of one-body systems, with the ensemble
then sampled by Monte Carlo (MC) techniques. We use open-ended random walks in Slater determinant space to sample the imaginary time projection and represent the ground state wave function: $|\psi_g \rangle
=\int d\phi\,c_\phi |\phi\rangle$, where the Slater determinants in the integral are
non-orthogonal $\langle \phi'|\phi\rangle\ne 0$.
A  gauge constraint, implemented approximately with a trial wave function  $|\psi_T\rangle$,
 is applied
on the sampled  Slater determinants  \cite{ShiweiConstraint, ShiweiPhaseless}
to control the sign or phase problem.

In this work, we present results obtained from the AFQMC method implemented for Gaussian basis sets \cite{Wissam-JCP124_224101_2006,Mario-WIRES-2018}.  We set the linear dependence threshold to be $10^{-8}$ for the one-electron basis \cite{Motta17} and use the modified Cholesky decomposition \cite{WirawanCholeskey}  with a threshold  $10^{-6}$ for the Coulomb interaction.  Most of the calculations use projection time $\beta = 35~E^{-1}_{Ha}$  and time step $\Delta \tau = 0.005~E^{-1}_{Ha} $. The convergence error from finite $\beta$ is negligible and extrapolations are performed when the Trotter error is larger than Monte Carlo uncertainty. The reported error bars  are estimated by one standard deviation statistical errors.

Truncated CASSCF wave functions were used as $|\psi_T\rangle$ here.
Fast update procedure \cite{HaoShiFastUpdate, doi:10.1021/acs.jctc.8b00342}
allows the use of multi-determinant CAS trial wave functions with sublinear cost.
 (For example, in ScO TZ basis, the cost of the AFQMC calculation with a $|\psi_T\rangle$  of $163$ determinants is $2.1\times $ that of a single determinant calculation.)
Typically around $10$ CAS orbitals are used to generate the $|\psi_T\rangle$
in both atoms and molecules.
 Following procedures in past AFQMC calculations using CASSCF  \cite{JChemPhys_127_144101_2007,WirawanCr2}, we truncate the wave function by discarding determinants with the smallest weights, up to an integrated weight of $\delta = 10^{-3}$, which results in $\sim 100$ determinants in most cases.
 Because of the fast update algorithm,
 we could check the effect on the AFQMC results
 of increasing the CAS space to the next level with little cost.  In VO and FeO, a noticeable difference was seen
 outside the statistical error, and we increased the CAS space to $12$, resulting in
 $\sim 1700$ and $\sim 1600$ determinants, respectively, in their $|\psi_T\rangle$.

In solids, CASSCF trial wave functions would not be applicable straightforwardly in a size-consistent
 manner.
There have been many benchmark studies of AFQMC using single determinant trial
wave functions (e.g. Refs.~\cite{JChemPhys_127_144101_2007,WirawanCr2, WirawanMolybdenum}).
To give an idea of the dependence of the constraint error in AFQMC for the specific systems here,
 the computed  total energies in  TiO, VO, CrO, MnO
 change by about
 $-4.9$, $-1.5$, $+6.8$, and $-9.7$ milli-Hartrees
 from the reported multi-det $|\psi_T\rangle$ results
 when a single determinant UHF trial wave function is used
 (TZ basis, with MC error bars $1$-$2$  milli-Hartrees).
 Besides using a single Slater determinant,
 there are a number of possibilities to systematically improve the trial wave function, including
 generalized Hartree-Fock \cite{MingpuBenchmark},
 symmetry projection and symmetry-adapted multi-determinants \cite{HaoShiSymmetryMD, HaoShiProjectedMD,Motta17},
  Hartree-Fock-Bogoliubov form \cite{FermiGasBCS, Hao-Shi-HFB}, self-consistent trial wave functions \cite{MingpuSelfConsistent, YuanyaoSelfConsistent}.

\subsection{Configuration interaction} 

Configuration interaction with singles and doubles excitations (CISD) was used as implemented in the PySCF package. 
CISD approximates the many-body wave function as a sum of Slater determinants, constructed from a reference Slater determinant, which was taken from restricted open shell Hartree-Fock. 
In CISD, the wave function is given as 
\begin{align}
	|\Psi_{CISD}\rangle = (&c_0 \notag \\
				     &+ \sum_{ij,\sigma} C^{(s)}_{ai} c_{a,\sigma}^\dagger c_{i,\sigma}\notag \\
				     &+
\sum_{ijkl,\sigma,\sigma'} C^{(d)}_{abij,\sigma,\sigma'} c_{a,\sigma}^\dagger c_{b,\sigma'}^\dagger  c_{i,\sigma} c_{j\sigma'} \notag \\ 
				     &) |\Psi_{HF}\rangle \notag,
\end{align}
where $c^\dagger$ and $c$ are creation/destruction operators, respectively, $a,b$ refer to virtual orbitals, $i,j$ refer to occupied orbitals, and all $C$ parameters are variationally optimized.
CISD scales approximately as $\mathcal{O}(N_e^6)$ and is known to not be size extensive.

\subsection{Coupled Cluster}

Unrestricted coupled cluster was used as implemented in the PySCF package.\cite{sun_pyscf}
The reference state was restricted open shell Hartree-Fock. 
We found that an unrestricted reference state led to worse results by values up to 10 mHa for the total energy of the molecules. 
In UCCSD, the wave function is approximated as 
\begin{equation}
	|\Psi_{CCSD}\rangle = e^{\hat T} |\Psi_{HF}\rangle,
\end{equation}
where the $\hat{T}$ operator contains one and two body operators. 
The exponential \textit{ansatz} ensures that the technique is size extensive in contrast to CISD. 
UCCSD scales approximately as $\mathcal{O}(N_e^6)$ \cite{bartlett_coupled-cluster_2007}.

UCCSD(T) evaluates the perturbative effect of including three-body operators from a UCCSD reference, and is often called the gold standard of quantum chemistry when used for equilibrium properties. It scales approximately as $\mathcal{O}(N_e^7)$. Despite the steep formal scaling, the prefactor is quite small.
Thus compared to the other accurate methods in this paper, namely, SHCI, DMRG, FCIQMC, AFQMC, and SEET, the UCCSD(T) calculations were
the least expensive by a significant amount.
%% and has been applied to extended systems, although typically with small basis sets on large systems \cite{bartlett_coupled-cluster_2007}.

\subsection{Density Functional Theory}

Density functional theory (DFT) in the restricted open shell Kohn-Sham approach was used as implemented in the PySCF package.\cite{sun_pyscf}
Level 6 grids were used to improve the accuracy, and the resultant state was carefully checked to ensure that it was the DFT ground state, since often the self consistent field process converged to the incorrect state. 
The basis set error in DFT is very small, typically with less than 1 mHartree difference between the vtz and v5z basis sets. 
This is because the basis in DFT only has to express the occupied Kohn-Sham orbitals accurately, and is not used to describe electron correlation. 
Strictly speaking, the DFT energy is only comparable to the many-body solution in the complete basis set, because the functionals are designed to approximate
  the correlation energy in the basis set limit.
%% since the DFT functionals are based on the $1/r$ Coulomb interaction, which is modified by a finite basis. 

\subsection{DMRG}

The density matrix renormalization group (DMRG)~\cite{white1992density} provides a variational ansatz for the wavefunction of the matrix product state form,
\begin{align}
  |\Psi\rangle = \sum_{n_1n_2 \ldots n_k} [\mathbf{A}^{n_1} \mathbf{A}^{n_2} \ldots \mathbf{A}^{n_k}]_{11} |n_1 n_2 \ldots n_k\rangle
\end{align}
where $\mathbf{A}^n$ is a matrix of variational parameters for each orbital and $|n_1 n_2 \ldots  n_k\rangle$ is an occupancy vector. The above
ansatz expresses the coefficient of any  occupancy vector as a product of matrices, where the ``bond'' dimension of
the matrix $\mathbf{A}^n$ is $M \times M$. $M$ may be increased until the ansatz is exact, which happens for $M$ approximately
the square root of the full Hilbert space size. The cost of the calculation using
the quantum chemistry Hamiltonian with quartic interactions is proportional to $M^3 k^3 + M^2 k^4$ where $k$ is the number of orbitals~\cite{white1999ab,chan2002highly,keller2016spin,chan2016matrix}.
In a localized basis, the $M$ required for a given accuracy scales like $e^{V^{D/D+1}}$ where $V$ is the volume of the system and $D$ is the dimension~\cite{hachmann2006multireference}.
Thus, when extending a system along one dimension, $M$ is independent of system size, while when extending a system along
all three dimensions, the computational scaling is $e^{V^{2/3}}$. In any dimension, this is therefore a savings over full configuration interaction,
which scales like $e^V$.

As can be seen, the ansatz requires an ordering of the orbitals and also treats all orbitals
on the same footing.  The latter means that in practice the DMRG is often a good ansatz relative to many methods when there are active orbitals
which needed to be treated in a balanced way. However, it is inefficient when there are many doubly occupied or empty orbitals. The
atoms and molecules in this system fall into this latter single-reference category. Thus we do not expect the DMRG to be
especially efficient, but it serves as a near-numerically exact method to benchmark other techniques more suitable for these systems.

The DMRG calculations in this work were carried out using a spin-adapted code (a slight modification of the above ansatz) which allows
us to obtain pure spin states~\cite{sharma2012spin}. The orbital ordering was generated by the default genetic algorithm~\cite{olivares2015ab}. 
We used the two-site variant of the DMRG and carried out calculations systematically increasing $M$.
The largest $M$ we used ranged from 4000 - 10000. To verify the accuracy of the energy
we carried out an extrapolation in the total energy. We did this either by the standard linear extrapolation in the energy against the discarded weight
in the two-site algorithm, where the DMRG energies at different $M$ were computed by sweeping backwards from the largest $M$ down to smaller $M$'s (backwards schedule)~\cite{legeza1996accuracy,chan2002highly,olivares2015ab}, or by extrapolating the energies of the largest $M$ values against $1/M$. In the first case, the extrapolation error
is usually reported as a fraction of the extrapolation distance between the lowest variational energy and the extrapolated energy. This extrapolated
energy was consistent with the energy obtained by extrapolating against $1/M$, but in some cases the $1/M$ extrapolation was more linear,
and we report that as the extrapolated energy. The DMRG energies at the largest $M$ are variational. Where included, they provide
the lowest variational energies for this benchmark.

\subsection{FCIQMC}
The FCIQMC method \cite{doi:10.1063/1.3193710,Booth2013,Booth2014} directly samples the many body wavefunction by stochastic propagation of a population of discrete walkers in Slater determinant space, defined by a given single particle basis. The annihilation of walkers with anti-walkers is crucial to ensure that the average wavefunction has the correct configurational sign-structure, and can lead to a circumvention of the Fermion sign problem without uncontrolled approximations. When a determinant is occupied by a small number of walkers, it is not clear whether the determinant should be ultimately dominated by walkers or anti-walkers, and so spawning new walkers from such a determinant can cause incorrect sign information to propagate throughout the network. This problem is minimized by the systematically improvable approach of Initiator FCIQMC \cite{doi:10.1063/1.3302277,bcta2011}, which only allows the creation of new walkers on currently unoccupied determinants, by parent walkers residing on a determinant with a population above some threshold (in this work taken to be 3 walkers). This increases the incidence of annihilation events and encourages the sign-coherent propagation of walkers, and results in a convergence to the exact wavefunction energy and properties as the number of walkers increases. Furthermore, small subspaces are identified to define a `trial wavefunction' onto which the sampled wavefunction is projected to calculate the energy, as well as another small subspace in which exact propagation can occur (the semi-stochastic adaptation \cite{doi:10.1063/1.4920975,Umrigar2012}). These subspaces serve to minimize the stochastic errorbars of the estimators.

All FCIQMC calculations in this work were undertaken via the following process:
\begin{itemize}
    \item A maximum walker population $N_\text{walker}$ is chosen
    \item The walker population is initialized on a single determinant and then allowed to reach $N_\text{walker}$
    \item A short interval in imaginary time after the maximum population is reached, the trial wavefunction space is initialized by performing an exact deterministic diagonalization of the Hamiltonian in the subspace spanned by the $N_\text{TWF}$ most populated determinants ($N_\text{TWF}\sim$200 determinants in this work)
    \item At the same iteration, a subspace of the $N_\text{SS}$ most populated determinants is identified and designated as the semi-stochastic space. Thereafter, the walkers residing in this subspace are exactly propagated. ($N_\text{SS}\sim$10,000 determinants in this work)
    \item The walker population is left to evolve under initiator FCIQMC dynamics until the numerator and denominator of the trial wavefunction projected energy stabilize around a mean value
    \item The energy is taken to be the ratio of means of the numerator and denominator of the energy
    \item The stochastic error is estimated using the Flyvberg -- Peterson analysis\cite{doi:10.1063/1.457480} for serially correlated data.
\end{itemize}

The stochastically sampled wavefunction is affected by the systematic error introduced by the initiator criterion for spawning. In order to reduce this error to within acceptable bounds, the above procedure is repeated for increasingly large values of $N_\text{walker}$ until convergence of the energy estimate with respect to this parameter is achieved. This was approximately 15 million, 50 million, 100 million and 200 million walkers for the Tm/Tm$^+$ vdz, vtz, vqz and TmO vdz calculations respectively\cite{tba2015}. This computational effort very roughly corresponds to 100 core hours per million walkers, with a maximum of $\sim$15,000 CPU hours used in order to converge to ScO (vdz) system to small random error bars (200 million walkers).

\subsection{Fixed node diffusion Monte Carlo} 

Fixed node diffusion Monte Carlo was used as implemented in the QWalk package.\cite{wagner_qwalk:_2009}
A single determinant was generated using PySCF density functional theory in the B3LYP approximation.
This determinant gave the lowest upper bound energy to the ground state. 
We multiplied the determinant by a 3-body Jastrow factor, which then was optimized to minimize the total energy using the linear method.\cite{toulouse_2007,toulouse_full_2008}
The resultant single determinant Slater-Jastrow wave function was used as a guiding function for diffusion Monte Carlo.\cite{foulkes_quantum_2001}
In this method, the diffusion Monte Carlo wave function is given by 
\begin{equation}
	|\Psi_{DMC}\rangle = e^{-\tau \hat{H}} |\Psi_T\rangle,
\end{equation}
where $|\Psi_T\rangle$ is the trial wave function, in this case the Slater-Jastrow wave function.
The T-moves scheme\cite{casula_beyond_2006} was used to ensure an upper bound to the ground state energy.
Timestep errors were extrapolated out using a linear fitting process and timesteps as low as 0.0025 Hartrees$^{-1}$. 

Scaling in stochastic methods is complex. 
To obtain the total energy with a given stochastic uncertainty, DMC(SD) scales as $\mathcal{O}(N_e^3+\epsilon N_e^4)$.
The $N_e^4$ cost typically does not appear until the number of electrons is more than 400-500. 
This method has been applied on systems with more than 1000 electrons.

We also considered more accurate trial wave functions, which can improve the fixed node error. 
We constructed them from SHCI wave functions by running small selected CI, and choosing the determinants with the largest weights. 
These weights were then reoptimized in the presence of the Jastrow factor using the linear method.\cite{toulouse_2007,toulouse_full_2008}

\subsection{GF2}

The fully self-consistent second order Green's function theory (GF2) \cite{Phillips14,Rusakov16,Phillips15,Welden16,Dahlen04b} includes all second-order skeleton diagrams dressed with the renormalized second-order propagators and bare interactions. GF2 is formulated as a low-order approximation to the exact Luttinger-Ward (LW) functional \cite{Luttinger60}  and therefore is $\Phi$-derivable, thermodynamically consistent, and conserving \cite{Baym61,Baym62}.

For transition atoms, we solve all the non-linear equations self-consistently at non-zero temperature. At each iteration, the self-energy, Green's function, and Fock matrix are updated until convergence is reached, so that the converged solution is reference-independent. Because of stability problems in the self-consistency, the calculations for transition monoxides are only done with one-shot GF2 on top of unrestricted Hartree-Fock. The self-energy, Green's function, and Fock matrix are not iterated until self-consistency. All the calculations are done on a mesh that combines a sparse power-law grid with an explicit transform based on a Legendre expansion of the self-energy. \cite{Kananenka16}

\subsection{MRLCC}

\newcommand{\bra}[1]{\ensuremath{\langle #1 \vert}}
\newcommand{\ket}[1]{\ensuremath{\vert #1  \rangle}}
\newcommand{\expect}[3]{\ensuremath{\langle  #1 \vert #2 \vert #3 \rangle}}
The multi-reference linearized coupled-cluster (MRLCC) is a flavor of multi-reference perturbation theory.
We consider a reference wavefunction $\ket{\Psi_0}$ obtained for example from a CAS-like calculation.
The out-of-active-space dynamical correlation may
be added by multi-reference perturbation theory,
where the expressions of the four first contributions to
the energy are
\begin{align}
E_0&=\expect{\Psi_0}{\hat{H}_0}{\Psi_0}
\label{eq:E0}
\\
E_1&=\expect{\Psi_0}{\hat{V}}{\Psi_0}
\label{eq:E1}
\\
E_2&=\expect{\Psi_0}{\hat{V}}{\Psi_1}
\label{eq:E2}
\\
E_3&=\expect{\Psi_1}{\hat{V}-E_1}{\Psi_1}
\label{eq:E3}
\end{align}
and where the first order correction to the wavefunction $\ket{\Psi_1}$ obeys
\begin{align}\label{eq:Psi1}
(E_0-\hat{H}_0)\ket{\Psi_1}=\hat{V}\ket{\Psi_0}
.
\end{align}

In Rayleigh-Schr\"{o}dinger perturbation theory, the partitioning of the Hamiltonian
$\hat{H}=\hat{H}_0+\hat{V}$ is so that $\ket{\Psi_0}$ and $E_0$ are the
eigenvector and eigenvalue of the zeroth order Hamiltonian $\hat{H}_0$. 
As is well known, this leaves some flexibility for the choice of $\hat{H}_0$,
leading to different perturbation theories having different properties:
the use of the Fock operator yields CASPT\cite{caspt},
and the use of the Dyall Hamiltonian\cite{dyall} yields NEVPT\cite{nevpt1,nevpt2}.
In this work,
we show derivations and results for the MRLCC perturbation theory\cite{mrlcc1,mrlcc2,mrlcc3,mrlcc4},
which uses the Fink Hamiltonian\cite{fink1,fink2}:
\begin{align}\label{eq:H0}
    \hat{H}_0=\Big(\sum_{mn}t_m^n \hat{E}_m^n + \sum_{mnop}v_{mn}^{op} \hat{E}_{mn}^{op}\Big)_{\Delta=0}
,
\end{align}
where the spin-free excitation operators are written with a hat, $t$ and $v$ are tensors, and the $m,n,o,p$ indices refer to any molecular orbital.
The notation $\Delta=0$ indicates that only terms that do not change the number of electrons
in the core, active and virtual spaces are taken into $\hat{H}_0$.
It follows that:
\begin{align}\label{eq:V}
    \hat{V}=\Big(\sum_{mn}t_m^n \hat{E}_m^n + \sum_{mnop}v_{mn}^{op} \hat{E}_{mn}^{op}\Big)_{\Delta\neq 0}
,
\end{align}
and one can readily see from Eqs.~(\ref{eq:E0}) and (\ref{eq:E1})
that the zeroth order energy is the energy of the reference wavefunction and that $E_1=0$,
which goes to say that this zeroth order Hamiltonian 
is somewhat close to the exact $\hat{H}$
and is good in the context of perturbation theory
(this is also the case for NEVPT).

In the internally contracted scheme,
the first order correction to the wavefunction is expressed
as a sum over eight class contributions $\ket{\Psi_1^c}$
expanded on perturber wavefunctions that are connected to the reference wavefunction:
\begin{align}\label{eq:span}
\ket{\Psi_1^c} = \sum_I d^c_I \,\,\hat{E}_I^c\ket{\Psi_0}
,
\end{align}
with coefficients $\mathbf{d}^c$.
In this representation the Fink Hamiltonian is block diagonal (and so is the Dyall Hamiltonian),
and the coefficients $\mathbf{d}^c$ in Eq.~(\ref{eq:span})
are found by solving Eq.~(\ref{eq:Psi1}) subsequently for each class.
Projection of  Eq.~(\ref{eq:Psi1}) onto the basis of a class yields:
\begin{align}
\label{eq:AdSw}
\mathbf{A}^c\mathbf{d}^c =\mathbf{S}^c\mathbf{w}^c
,
\end{align}
with
\begin{align}
\label{eq:A}
& A^c_{IJ}
=\expect{\Psi_0}{\hat{E}^c_I{}^\dagger\left[(E_0-\hat{H}_0),\hat{E}^c_J\right]}{\Psi_0}
\\
\label{eq:S}
& S^c_{IJ}
=\expect{\Psi_0}{\hat{E}^c_I{}^\dagger                      \hat{E}^c_J       }{\Psi_0}
\end{align}
where realizing that $(E_0-\hat{H}_0)\ket{\Psi_0}=0$ allows the introduction of the commutator
(see for example Ref.~\citenum{mrlcc4}).

%The tedious part in terms of computation is that
%the perturber wavefunctions used as basis in the internally contracted scheme
%are in general non-orthogonal many-body states
%and Eq.~(\ref{eq:AdSw}) involves overlaps.
%In practice, this is tackled using an orthonormalization procedure\cite{mrlcc}.

The terms to manipulate to compute $\mathbf{A}$, $\mathbf{S}$ and $E_2$, $E_3$ 
involve long strings of creation/annihilation operators,
and one can use the Wick's theorem to simplify the expressions.
This will result in series of tensor contractions involving
one and two electron integrals (stemming from $\hat{H}_0$ and $\hat{V}$),
and RDMs up to fourth order.
The application of the Wick's theorem to the strings of operators
is done with the ``Second Quantization Algebra'' symbolic algebra Python library~\cite{neuscamman2009quadratic},
which we modified to fit our needs.
The scripts also automate the generation of the C code used to solve the resulting equations
and to calculate the energies $E_2$ and $E_3$.

\subsection{QSGW}

Typically RPA total energies are done from a simple one-body reference Hamiltonian $H_{0}$ such as PBE, HSE06, or
Hartree-Fock.  Since different choices yield different results, there are significant and unavoidable ambiguities.
First of all, there is a fundamental issue: When employing a non-self-consistent Green's function, the different
formulas for the total energy do differ in practice: the Galitskii-Migal and the RPA total energies are not equal any
more \cite{dahlen_pra2006}.

The starting-point dependence can be surmounted by iterating $G$ to self-consistency, that is, by finding a $G$
generated by \emph{GW} that is the same as the $G$ that generates it ($G^\mathrm{out}$=$G^\mathrm{in}$).  But it has
long been known that full self-consistency can be quite poor in solids \cite{Shirley96,holm98}.  A recent re-examination
of some semiconductors \cite{Grumet18} confirms that the dielectric function (and concomitant QP levels) indeed worsen
when $G$ is self-consistent, for reasons explained in Appendix A in Ref.~\cite{Kotani07}.  Fully sc$GW$ becomes more
problematic in transition metals \cite{BelashchenkoLocalGW}.  Finally, sc\emph{GW} is a conserving approximation in the
Green's function $G$, but $W$ loses its usual physical meaning as a response function.

An alternative is the Quasiparticle Self-Consistent \emph{GW} approximation\cite{Faleev04} (QS\emph{GW}).  It is similar
to sc$GW$, but at each cycle the dynamical self-energy is rendered static and hermitian, forming a new noninteracting
$G_{0}$ by making the substitution
\begin{eqnarray}
V^{\rm xc} = \frac{1}{2}\sum_{ij} |\psi_i\rangle 
       \left\{ {{\rm Re}[\Sigma(\varepsilon_i)]_{ij}+{\rm Re}[\Sigma(\varepsilon_j)]_{ij}} \right\}
       \langle\psi_j|,
\label{eq:veff}
\end{eqnarray}
where $\mathrm{Re} \{ \}$ stands for the Hermitian part of the operator.
This process is carried through to self-consistency.  It was formally justified \cite{Faleev04,Kotani07} as a
construction to optimize the noninteracting Green's function $G_{0}$ by minimizing some measure of the difference
$|G-G_{0}|$.  More recently it has been justified as minimizing the the gradient of the Baym-Kadanoff functional in the
space of all $G_{0}$ \cite{Beigi17}.

QS\emph{GW} is nevertheless an approximate self-consistent procedure, which relies on effective one-electron
wavefunctions instead of the full Green's function.  As a consequence, the difference between the expressions for the
total energy (Galitskii-Migdal or RPA) still persists once the self-consistency has been reached.  In this work, we
defined the QS\emph{GW} total energy as the one obtained from the RPA expression based on QS\emph{GW} eigenvalues
and wavefunctions, consistently with Ref.~\onlinecite{bruneval_jcp2012}.  Indeed, the RPA total energy, through the
adiabatic connection, is capable of incorporating the correlated part of the kinetic energy \cite{fuchs_jcp2005},
whereas the Galitskii-Migdal is not. The correlated part of the kinetic energy is sizeable and it is
important to include it properly.

It is well known that \emph{GW} overestimates the correlation energy.  A prior study of weakly correlated molecular
dimers\cite{OlsenRPA+ALDAEnergy} showed that (1) the RPA tends to systematically overestimate the correlation energy,
and (2) the error is connected with short-ranged correlations.  This tendency is also found here, as noted in the main
text.  The ionization energy and the dimer formation energy, both of which benefit from partial cancellation of errors
in short-range correlation, are much better described.  We also find that the RPA total energy based on QS\emph{GW},
with its optimal choice for $G_{0}$, perform significantly better than RPA based on other $G_{0}$, e.g. PBE or Hartree
Fock, as will be shown elsewhere.

Finally, it is has been established, using less optimal forms for $G_{0}$, that low-order diagrammatic corrections
(especially second order screened exchange) significantly reduce errors in the correlation energy
\cite{ren_prb2015,maggio_jctc2017}.  It is shown elsewhere \cite{Cunningham18} how ladders dramatically improve the
dielectric function in TM oxide crystals, so it is reasonable to expect that correlation energies computed from it will
see a similar improvement.  Density-functional approximations for the exchange-correlation kernel significantly improve
on heats of formation of dimers from \emph{sp} elements~\cite{PhysRevMaterials.2.075003}.

\subsection{RPA}

We calculated the RPA total energy using the Tamm-Dancoff approximation (specifically Eq. (9) from
Ref.~\cite{Eshuisa10}), using several flavors of $G_{0}$: PBE, Hartree-Fock, HSE06, and QS\emph{GW}, for the TM atom and
M+O dimer.  
All the RPA calculations in this paper use the MOLGW code~\cite{Bruneval16}.

\subsection{sc-GW}
The $GW$ method \cite{Hedin65} evaluates a subset of terms of a diagrammatic weak coupling series in the interaction $V$ deterministically. The $GW$ approximation can be understood as a first-order approximation to Hedin's series of renormalized propagators and interactions. It is expressed in terms of self-energies $\Sigma$, Green's functions $G$, screened interactions $W$ and polarizations $P$ by the self-consistent solution of the equations $G=G_0+G_0\Sigma G$, $W=V+VPW$, where $G_0$ denotes the Hartree-Fock Green's function, and $\Sigma$ and $P$ are computed as  $\Sigma=-GW$ and $P=GG$. The main difference to GF2 is that, in $GW$, both propagators and interactions are renormalized, whereas GF2 only renormalizes the propagators. However, GF2 obtains all second order contributions, whereas the second order exchange is missing from $GW$. Our results are converged to self-consistency, using a finite-temperature imaginary time formulation evaluated at temperatures low enough that the system is in its ground state. Sparse imaginary time and Matsubara frequency grids based on Chebyshev polynomials and the intermediate representation (IR) \cite{Gull18,shinaoka2017compressing,li2019sparse} is used, which significantly reduced the computational cost. Our code is based on the ALPS libraries \cite{Gaenko17,Wallerberger18}.

Self-consistent GW is a $\Phi-$ \cite{Baym61,Baym62} and $\Psi$-derivable \cite{Almbladh99} weak coupling method, in the sense that it neglects some diagrams of order $V^2$. Achieving full self-consistency requires the storage and manipulation of $W$, which is a frequency-dependent four-index tensor. The necessity of handling this object numerically restricts the method in our implementation to relatively small system sizes.

\subsection{Self-energy embedding theory (SEET)}
SEET~\cite{zgid_njp17,Zgid15,Tran15b,Tran16,GWSEET2017,Tran_generalized_seet,Tran_useet,seet_periodic1} is a finite temperature Green's function embedding method. 
The embedding construction allows us to describe the weakly and strongly
correlated orbitals at different levels of theory. The weakly correlated orbitals
are treated by a low level, most often a perturbative method (here Green's function second order (GF2)~\cite{Zgid14,Phillips15,Rusakov16,Welden16,Neuhauser16,kananenka_hybrif_gf2}	
or a single iteration of GF2). The strongly correlated orbitals are treated with a high level,
usually non-perturbative method. When multiple strongly correlated orbitals are present, they are
separated into several intersecting or non intersecting subsets $A_i$. Each of these subsets contains $M_i^A$ orbitals and $M=\sum_i M_i^A +M^R$, where $M$ is the total number of orbitals in the problem and $M^R$ are all the orbitals that are not contained in the groups of the strongly correlated orbitals.
The orbitals from each of the subsets $A_i$ are used to construct Anderson Impurity Models (AIM) that are then solved by a non-perturbative method,
here full configuration interaction (FCI)~\cite{Zgid11,ZgidPRB2012,PhysRevB.96.235149}. The intersubset interactions are treated most commonly at a perturbative level.
Orbitals chosen to each of the $A_i$ subsets can be chosen based on several criteria such as occupancies of natural orbitals (NOs)
or energies of molecular orbitals (MOs), for details see Refs. [48,50].

A general SEET functional can be written as
\begin{eqnarray}
\Phi^\mathrm{SEET}_\mathrm{MIX}& =\Phi^\mathrm{tot}_\mathrm{weak} +
\sum_{i}^{n \choose k} (\Phi^\mathrm{A_i^k}_\mathrm{strong}-\Phi^\mathrm{A_i^k}_\mathrm{weak})\\ \nonumber
&\pm \sum_{k=K-1}^{k=1}\sum_{i}^{n \choose k} (\Phi^\mathrm{B_i^k}_\mathrm{strong}-\Phi^\mathrm{B_i^k}_\mathrm{weak}),
\end{eqnarray}
where $\Phi^\mathrm{tot}_\mathrm{weak}$, in this work is a GF2 solution for the whole orbital space,
$\Phi^\mathrm{A_i^k}_\mathrm{strong}$ is obtained from the solution of AIM for the strongly correlated subset of orbitals
$A_i$, $\Phi^\mathrm{A_i^k}_\mathrm{weak}$  is the solution of the subset $A_i$ with a weakly correlated method
used to remove the double counting. The terms $(\Phi^\mathrm{B_i^k}_\mathrm{strong}-\Phi^\mathrm{B_i^k}_\mathrm{weak})$
are present in case of intersecting subsets $A_i$ and are necessary to remove the double counting, for details see Ref.~\onlinecite{Tran_generalized_seet}.
We denote a particular SEET calculation	as SEET(method strong/method weak)-m($\mathrm{[M^Ao]}$/basis) since self-energies from intersecting
orbital subspaces containing $M^A$ orbitals are	treated	with ``method	strong".	The whole system is treated with ``method weak" and the orbitals from
subsets	$A_i$ are transformed to a certain orbital basis denoted here as ``basis". In this paper, we most commonly use the basis	of molecular orbitals.	
The details of the finite temperature imaginary time GF2 grid as well as the frequency grid can be found in Ref.~\onlinecite{Iskakov_Chebychev_2018,Kananenka16,Kananenka15}.

\subsection{SHCI} 
%In this section, we review the semistochastic heat-bath configuration interaction method (SHCI)~\cite{HolTubUmr-JCTC-16,ShaHolJeaAlaUmr-JCTC-17,HolUmrSha-JCP-17},
%emphasizing the two important ways it differs from other SCI+PT methods.
%%In common with other SCI+PT methods, SHCI has two stages, the variational stage and the perturbation stage.
%%{\color{red} verbatim repetition from intro:} It differs from other SCI+PT methods in three important ways.
%%First, it uses a different and much faster determinant selection criterion~\cite{HolTubUmr-JCTC-16}.
%%%Second, it uses ``auxiliary arrays" to speed up the calculation of the Hamiltonian matrix~\cite{ShaHolJeaAlaUmr-JCTC-17}.
%%Second, it uses an efficient semistochastic algorithm to compute the perturbative correction that overcomes
%%the memory bottleneck in that stage of the calculation~\cite{ShaHolJeaAlaUmr-JCTC-17}.
%%(A different semistochastic algorithm has been proposed by Garniron et al.~\cite{GarSceLooCaf-JCP-17}, which is also efficient.)
%In the following, we use $\V$ for the set of variational determinants, and $\P$ for the set of perturbative determinants, that is, the set of determinants that are connected to the variational determinants by at least one non-zero Hamiltonian matrix element but are not present in $\V$.

The semistochastic heat-bath configuration iteration (SHCI) method~\cite{HolTubUmr-JCTC-16,ShaHolJeaAlaUmr-JCTC-17,HolUmrSha-JCP-17,LiOttHolShaUmr-JCP-18},
is an efficient instance of the general class of methods
wherein a selected configuration interaction is performed followed by a perturbative correction (SCI+PT).
SCI+PT methods have two stages.
In the first stage a set of ``important" determinants are selected the Hamiltonian is diagonalized in the subspace of these
determinants, $\mathcal{V}$, to obtain the
%$\mu^{th}$ state $|\Psi^{\mu}\rangle = \sum_{D_i \in \mathcal{V}} c_i^{\mu} |D_i\rangle$ with energy $E^{\mu}$.
the lowest few eigenstates (or the lowest state if one is interested in the ground state only).
%The wavefunction of the $\mu^{th}$ state is $|\Psi^{\mu}\rangle = \sum_{D_i \in \mathcal{V}} c_i^{\mu} |D_i\rangle$ and has energy $E^{\mu}$.
In the second stage, a second-order perturbation theory is used to calculate the energy contributions of all determinants that do not belong
to the space $\mathcal{V}$ but have a non-zero Hamiltonian matrix element with at least one of the determinants in $\mathcal{V}$.
Such methods have been used for about 50 years~\cite{BenDav-PR-69,HurMalRan-JCP-73,BuePey-TCA-74}
and continue to be a subject of interest~\cite{Eva-JCP-14,LiuHof-JCTC-16,SceAppGinCaf-JCoC-16,TubLeeTakHeaWha-JCP-16,GarSceLooCaf-JCP-17,DasMorSceFil-JCTC-18,GarSceGinCafLoo-JCP-18,LooSceBloGarCafJac-JCTC-18,HaiTubLevWhaHea-JCTC-19} to the present day.
%The SHCI algorithm improves upon other SCI+PT methods, first by reducing the computational time of performing both the variational and
%perturbative correction, and, second by eliminating the memory bottleneck for the perturbative calculation.
%The SHCI algorithm is much more time and memory efficient than other SCI+PT methods. First it reduces the computational time of performing both the variational and
%perturbative correction, and, second it eliminates the memory bottleneck for the perturbative calculation.

We briefly describe the two innovations that account for the time- and memory-efficiency of SHCI.
A more detailed description can be found in \onlinecite{LiOttHolShaUmr-JCP-18}.
\begin{enumerate} %[leftmargin=*,labelindent=0mm,labelsep=2mm,noitemsep,nosep]
\item During the variational and the perturbative steps, straightforward implementations of SCI+PT scan
all determinants connected to at least one of the
determinants in $\mathcal{V}$ and select those determinants ($|D_a\rangle$) for which the absolute value of the
$2^{nd}$-order perturbative contribution to the energy
\begin{align}
\left|\frac{\left(\sum_{D_i \in \mathcal{V}} H_{ai} c_i\right)^2}{E_V - E_a} \right| > \epsilon,
\end{align}
where the subscript $a$ denotes a determinant not currently present in $\mathcal{V}$,
$E_V$ is the energy of the current variational wavefunction, $E_a$ is the energy of determinant $D_a$, and $\epsilon$ is
a parameter that controls the number of determinants selected.
%Some more recent SCI+PT methods avoid scanning all connected determinants, but all of them scan many determinants
%that do not contribute significantly to the exact wave function.
During the variational stage, this is done iteratively to build up the variational wavefunction starting from
a single determinant.
During the perturbative stage, $\epsilon=0$.
Instead, SHCI modifies the selection criterion to~\cite{HolTubUmr-JCTC-16},
\begin{align}
\max_{D_i \in \mathcal{V}}\left|  H_{ai} c_i \right| > \epsilon,
\label{eq:hci}
\end{align}
which greatly reduces the cost by taking advantage of the fact that most of the matrix elements, $H_{ai}$, are 2-body excitations, which
depend only on the indices of the 4 orbitals whose occupations change and not on the other occupied orbitals of a determinant. Thus by
presorting the absolute values of all possible matrix elements of the 2-body excitations in descending order, the scan over determinants $D_a$
can be terminated when $|H_{ai}|$ drops below $\epsilon/c_i$.
A  similar idea is used to speed up the selection of 1-body excitations as well.
This enables a procedure in which
%\emph{only the important determinants which will contribute to the computed variational or the perturbative energy are ever looked at},
\emph{only the important determinants which will be included in the variational wavefunction, or make significant contributions to the perturbative correction, are ever looked at},
resulting in orders of magnitude saving over a naive implementation of the SCI+PT algorithm!
Different values of $\epsilon$ are used during the variational and the perturbative stages of the calculation,
which we denote by $\epsilon_1$ and $\epsilon_2$.
\item The first innovation greatly speeds up both the variational and the perturbative steps of the algorithm.
However, the perturbative step has a very large memory requirement when $\V$ has a large number of
determinants (say $10^9$) because all determinants that are connected
to those in $\V$ must be stored.
\footnote{An alternative straightforward approach does not have a large memory requirement, but
requires considerably larger computation time.}
We have developed a 2-step~\cite{ShaHolJeaAlaUmr-JCTC-17} and later a 3-step~\cite{LiOttHolShaUmr-JCP-18} semistochastic perturbative approach that both
completely overcomes this memory bottleneck and is faster than the deterministic approach.
\end{enumerate}

In addition to these two major methodological improvements, the SHCI method uses auxiliary arrays to speed up the computation of the
Hamiltonian matrix~\cite{LiOttHolShaUmr-JCP-18}.  Further, it makes extensive use of hashing and techniques such as variable-byte encoding, hardware atomic operations,
dynamic load-balancing and thread pooling to achieve a high efficiency in the use of computer time and memory.

The convergence of the variational and perturbative energies depends significantly on the orbitals used.
The convergence obtained from using Hartree-Fock orbitals can be improved by using natural orbitals obtained
from an SHCI calculation with a fairly large value of $\epsilon_1$, and can be further improved by using
orbitals that minimize the SHCI variational energy using a modified version of the algorithm described
in Ref.~\onlinecite{SmiMusHolSha-JCTC-17}.

We typically choose $\epsilon_2 = 10^{-6} \epsilon_1$, so that a single parameter, $\epsilon_1$ controls
the accuracy of the calculation.
%$\epsilon_1$ a systematic convergence to the full configuration interaction limit is obtained.
The energy at the $\epsilon_1=0$ limit is obtained using a quadratic fit to the energies
versus the perturbative correction~\cite{HolUmrSha-JCP-17}.
Note that although the SHCI algorithm has a perturbative component,
systematically improvable approximations to the exact energy in the chosen basis are obtained by performing calculations with
progressively smaller values of $\epsilon_1$ until the total energy (variational energy plus
perturbative correction), or its extrapolation versus the perturbative correction, is converged to the
desired tolerance.

\section{Basis sets, bond lengths, and effective core potentials} 

%All calculations were computed using the correlated gaussian basis sets and effective core potentials produced by J. Trail and Needs\cite{trail_correlated_2015,trail_pseudopotentials_2013}
All calculations used the effective core potentials and associated aug-ccpVnZ gaussian basis sets of Trail and Needs\cite{trail_pseudopotentials_2013,trail_correlated_2015,trail_shape_2017}.
%These effective core potentials are produced from explicitly correlated multi-configuration Hartree-Fock calculations, and take core-core and core-valence effects into account. 
These effective core potentials are produced from explicitly correlated multi-configuration Hartree-Fock calculations, and include contributions from core-core and core-valence correlations.
%Basis sets of double-, triple-, quadruple-, and quintuple-zeta levels of quality were used. 
Augmented double-, triple-, quadruple-, and quintuple-zeta basis sets were used. 

The molecules were computed with bond lengths in {\AA} as follows: 
ScO: 1.668, TiO: 1.623,VO: 1.591, CrO: 1.621, MnO: 1.648, FeO: 1.616, CuO: 1.725.
\section{Data}

Each contributing author provided a separate data file detailing the results of their calculations for both atomic and molecular systems. 
%Each row of each data file indicates a system considered, the method used, the basis set and effective core potential used, and the resulting total energies respectively, along with any associated systematic or stochastic error. 
Each row of each data file indicates the system considered, the method and the basis set used, and the resulting total energies, along with any associated systematic or stochastic error. 
Additional method-specific information is also included (for example, the size of the active orbital space used in auxiliary-field quantum Monte Carlo).
Not all methods completed each calculation using each level of basis set quality, due to the high computational expense of some methods.

The CSV headers required are: 
\begin{itemize}
    \item charge: charge of the molecule/atom (either 0 or 1)
    \item molecule: (for molecules) the name of the molecule: example ``VO''
    \item atom: (for atoms) the name of the atom: example ``V''
    \item pseudopotential: always ``trail'' for this test set
    \item pyscf-version: ``new'' for after pyscf 1.5. A small improvement in accuracy was implemented after this version. Should be ``new'' for all calculations
    \item method: a string representing the method used for the calculation. example ``PBE''
    \item totalenergy: total energy in Hartrees
    \item totalenergy-stocherr : Stochastic error estimate 
    \item totalenergy-syserr : Systematic error estimate, if available
\end{itemize}
%{\color{red} Since all calculations used the ``new'' version of PySCF, is it still worthwhile to comment on this?  May be yes, just in case some readers used older versions.}

A script called 'gather.py' retrieves all data from all directories. 
Further scripts handle plotting and error estimation.

\section{Energy comparison for several accurate methods} 
There are 5 methods (DMRG, iFCIQMC, UCCSD(T), AFQMC(MD) and SEET(FCI/GF2)) for which the
total energies have an rms deviation of 4 mHa or less relative to the SHCI reference.
The maximum absolute error, the rms error and the number of systems treated
are shown in Table~\ref{tab:te_error} for these methods.
Tables \ref{tab:ip_error} and \ref{tab:be_error} show the corresponding quantities
for the ionization energy and the dissociation energy.

\begin{table}[htb]
\label{tab:te_error}
\caption{Total energy errors relative to SHCI of the five methods that agree best with SHCI.
   }
\begin{ruledtabular}
\begin{tabular}{lrrr}
Method & \# systems & max abs error & rms error \\
\hline
DMRG &   39 &  0.001010 &  0.000238 \\
iFCIQMC &   49 &  0.001611 &  0.000639 \\
UCCSD(T) &   92 &  0.006811 &  0.002309 \\
%AFQMC(MD) &   88 &  0.007282 &  0.003343 \\ % omitting Cu+
AFQMC(MD) &   92 &  0.007470 &  0.003540 \\
SEET(FCI/GF2) &   59 &  0.013656 &  0.004001 \\
\end{tabular}
\end{ruledtabular}
\end{table}

\begin{table}[htb]
\label{tab:ip_error}
\caption{Ionization energy errors relative to SHCI of the five methods that agree best with SHCI.
   }
\begin{ruledtabular}
\begin{tabular}{lrrr}
Method & \# systems & max abs error & rms error \\
\hline
DMRG &   14 &  0.000826 &  0.000307 \\
iFCIQMC &   21 &  0.000965 &  0.000567 \\
UCCSD(T) &   28 &  0.001222 &  0.000675 \\
%AFQMC(MD) &   24 &  0.001580 &  0.000838 \\ $ omitting Cu+
AFQMC(MD) &   28 &  0.008400 &  0.002888 \\
SEET(FCI/GF2) &   18 &  0.006580 &  0.002466 \\
\end{tabular}
\end{ruledtabular}
\end{table}

\begin{table}[htb]
\label{tab:be_error}
\caption{Dissociation energy errors relative to SHCI of the five methods that agree best with SHCI.
   }
\begin{ruledtabular}
\begin{tabular}{lrrr}
Method & \# systems & max abs error & rms error \\
\hline
DMRG &    7 &  0.000678 &  0.000327 \\
iFCIQMC &    1 &  0.000488 &  0.000488 \\
UCCSD(T) &   28 &  0.005108 &  0.002990 \\
AFQMC(MD) &   28 &  0.007880 &  0.002590 \\
SEET(FCI/GF2) &   14 &  0.008356 &  0.004152 \\
\end{tabular}
\end{ruledtabular}
\end{table}

\section{Basis set extrapolation}

Methods not operating directly in the complete basis set limit must be extrapolated using
finite basis calculations.
The extrapolated energy $E_m(\text{CBS})$ for method $m$ is estimated from
\begin{equation}
E_m(\text{CBS}) = E_{\text{HF}} (\text{CBS}) + \Delta({\text{CBS}}).
\end{equation}
where $E_{\text{HF}} (\text{CBS})$ is obtained by fitting the HF energies to the form
\begin{equation}
E_{\text{HF}}(n) = E_{\text{HF}} (\text{CBS} ) + b \exp(-cn).
\end{equation}
and $\Delta({\text{CBS}})$ is obtained by fitting the correlation energies to
\begin{equation}
E_{m} (n) - E_{\text{HF}} (n) = \Delta({\text{CBS}}) +\frac{\gamma}{n^3},
\end{equation}
where $n$ is the cardinal index of the basis.

\begin{figure}
    \includegraphics{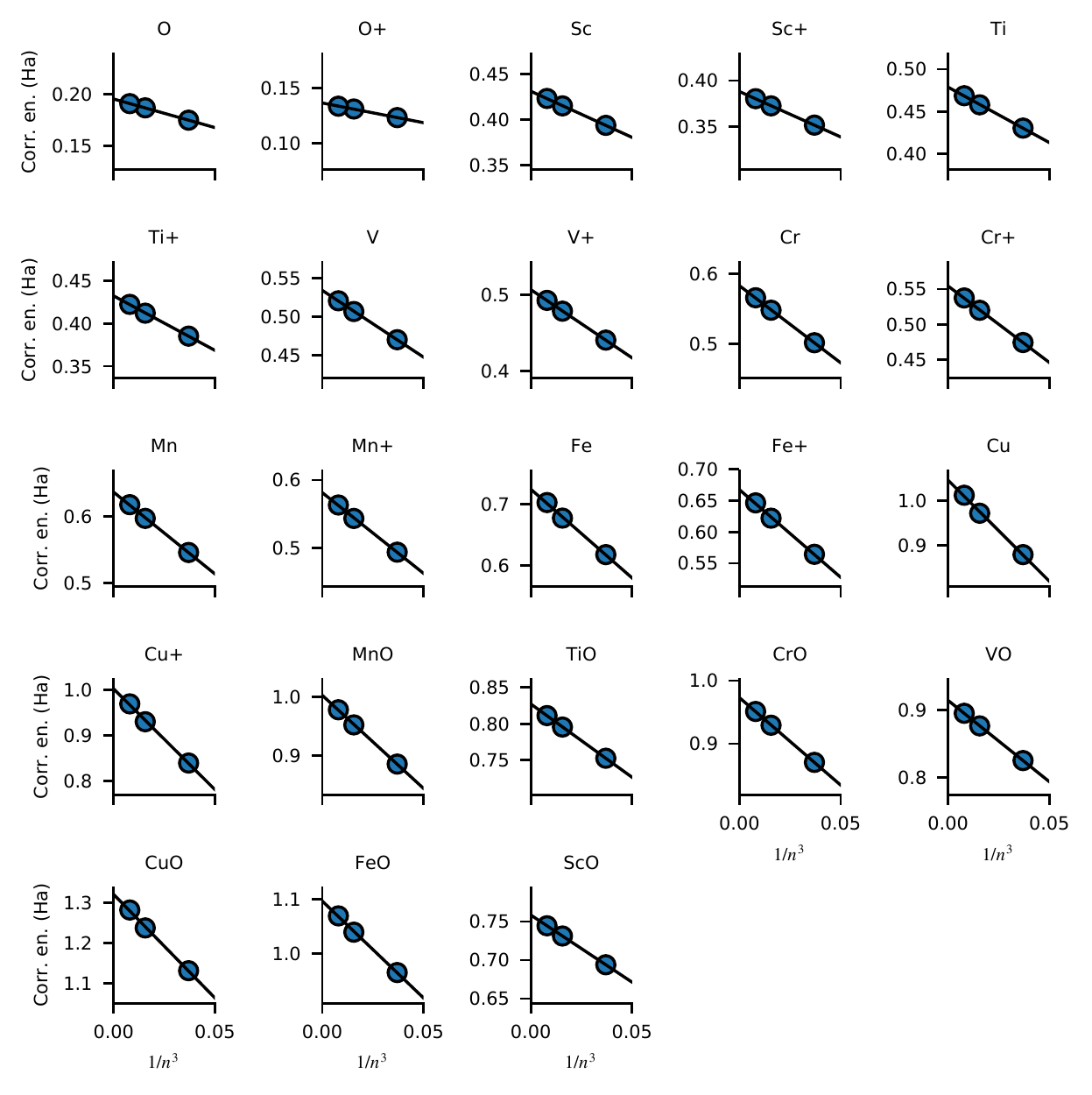} 
    \caption{Basis set extrapolation of the SHCI correlation energy versus the $n$ in the basis v$n$z. Only $n$ values from 3-5 are shown since the $n=2$ points deviate significantly from the straight lines shown.} 
    \label{fig:extrap}
\end{figure}

An example extrapolation for all materials and SHCI correlation energies is shown in Fig~\ref{fig:extrap}.
Extrapolations for the total energy, ionization energy and dissociation energy are shown in
%Tables~\ref{tab:te_basis_extrap, tab:ip_basis_extrap, tab:be_basis_extrap} respectively.
Tables~\ref{tab:te_basis_extrap}, \ref{tab:ip_basis_extrap} and \ref{tab:be_basis_extrap} respectively.
The ionization and dissociation energies converge much more rapidly than the total energies.
Of the 4 basis set extrapolations considered, cbs45 is the most accurate.
%Tables~\ref{tab:rms_ip_basis_extrap} and \ref{tab:rms_be_basis_extrap} show
Table~\ref{tab:rms_ip_be_basis_extrap} shows
the rms deviation
of the total, ionization and binding energies of each basis set and extrapolation
relative to the corresponding cbs45 energies.

\begin{table}
\caption{SHCI total energies for all basis sets and extrapolations.
   cbs23 indicates an extrapolation using only vdz and vtz bases, and cbs34/cbs45/cbs345 are labeled in the same way.
   %Uncertainties in the final extrapolated energies are of order 1-6 mHa.}
   The differences between the cbs345 and the cbs45 extrapolations increase with the atomic number of the
   transition metal and range from 1-8 mHa.  The cbs45 extrapolation is the most accurate one.}
   \label{tab:te_basis_extrap}
\begin{ruledtabular}
\begin{tabular}{lrrrrrrrr}
 & \multicolumn{8}{c}{Total Energy (Ha)} \\
 basis &          O &        Sc &        Ti &         V &        Cr &         Mn &         Fe &         Cu \\
 \hline
 vdz    &  -15.781250 & -46.396850 & -57.883050 & -71.077670 & -86.612030 & -103.942260 & -123.518120 & -197.236900 \\
 vtz    &  -15.827020 & -46.452300 & -57.954630 & -71.168980 & -86.720920 & -104.063660 & -123.657200 & -197.444640 \\
 vqz    &  -15.839090 & -46.474000 & -57.983000 & -71.205780 & -86.767610 & -104.115010 & -123.716750 & -197.536420 \\
 v5z    &  -15.843170 & -46.482130 & -57.993860 & -71.219830 & -86.785320 & -104.135420 & -123.741800 & -197.576080 \\
\hline
 cbs23  &  -15.846068 & -46.475646 & -57.985016 & -71.207033 & -86.766697 & -104.114189 & -123.715147 & -197.531655 \\
 cbs34  &  -15.847828 & -46.489701 & -58.003293 & -71.232489 & -86.801544 & -104.152337 & -123.760024 & -197.603262 \\
 cbs345 &  -15.847658 & -46.490126 & -58.004201 & -71.233375 & -86.802563 & -104.154264 & -123.763441 & -197.609436 \\
 cbs45  &  -15.847432 & -46.490691 & -58.005411 & -71.234555 & -86.803920 & -104.156833 & -123.767993 & -197.617661 \\
 \end{tabular}
\end{ruledtabular}

\begin{ruledtabular}
\begin{tabular}{lrrrrrrrr}
 & \multicolumn{8}{c}{Total Energy (Ha)} \\
 basis &          O+ &        Sc+ &        Ti+ &         V+ &        Cr+ &         Mn+ &         Fe+ &         Cu+ \\
 \hline
 vdz    &  -15.300670 & -46.156830 & -57.634020 & -70.828400 & -86.372410 & -103.673360 & -123.233540 & -196.967230 \\
 vtz    &  -15.333580 & -46.211490 & -57.704240 & -70.919550 & -86.472980 & -103.792140 & -123.368570 & -197.163040 \\
 vqz    &  -15.342100 & -46.233110 & -57.732560 & -70.957590 & -86.518670 & -103.842720 & -123.427090 & -197.253430 \\
 v5z    &  -15.344790 & -46.241150 & -57.743160 & -70.971820 & -86.536270 & -103.862840 & -123.451830 & -197.292510 \\
\hline
 cbs23  &  -15.348503 & -46.234171 & -57.733836 & -70.957555 & -86.515427 & -103.840708 & -123.424132 & -197.245135 \\
 cbs34  &  -15.348554 & -46.248705 & -57.752710 & -70.985120 & -86.552131 & -103.879380 & -123.469470 & -197.319302 \\
 cbs345 &  -15.348317 & -46.248980 & -57.753325 & -70.985640 & -86.553226 & -103.881194 & -123.472883 & -197.325395 \\
 cbs45  &  -15.348001 & -46.249346 & -57.754146 & -70.986333 & -86.554685 & -103.883611 & -123.477431 & -197.333514 \\
 \end{tabular}
\end{ruledtabular}

\begin{ruledtabular}
\begin{tabular}{lrrrrrrr}
 & \multicolumn{7}{c}{Total Energy (Ha)} \\
 basis &        ScO &       TiO &        VO &        CrO &        MnO &        FeO &        CuO \\
 \hline
 vdz    &  -62.420040 & -73.903750 & -87.085770 & -102.558370 & -119.850510 & -139.435990 & -213.123030 \\
 vtz    &  -62.530130 & -74.030850 & -87.235260 & -102.718670 & -120.029280 & -139.635460 & -213.377780 \\
 vqz    &  -62.568454 & -74.075488 & -87.288682 & -102.779756 & -120.097071 & -139.711352 & -213.484244 \\
 v5z    &  -62.582141 & -74.091473 & -87.307394 & -102.802737 & -120.122364 & -139.741682 & -213.529128 \\
 \noalign{\vskip 1mm}
 cbs23  &  -62.576098 & -74.084170 & -87.297330 & -102.786429 & -120.103370 & -139.717990 & -213.483952 \\
 cbs34  &  -62.595849 & -74.107270 & -87.326938 & -102.823579 & -120.145888 & -139.766095 & -213.561561 \\
 cbs345 &  -62.596114 & -74.107694 & -87.326918 & -102.824937 & -120.147133 & -139.769181 & -213.567787 \\
 cbs45  &  -62.596466 & -74.108258 & -87.326891 & -102.826745 & -120.148791 & -139.773293 & -213.576082 \\
 \end{tabular}
\end{ruledtabular}

\end{table}

\begin{table}
    \caption{SHCI ionization energies.
        cbs23 indicates an extrapolation using only vdz and vtz bases, and cbs34/cbs45/cbs345 are labeled in the same way.
        Experimental values are also shown.}
    \label{tab:ip_basis_extrap}
    \begin{ruledtabular}
    \begin{tabular}{llllllll}
     & \multicolumn{7}{c}{Ionization Potential (Ha)} \\
    basis &        Sc &        Ti &         V &        Cr &        Mn &        Fe &        Cu \\
    \hline
    vdz    &  0.240020 &  0.249030 &  0.249270 &  0.239620 &  0.268900 &  0.284580 &  0.269670 \\
    vtz    &  0.240810 &  0.250390 &  0.249430 &  0.247940 &  0.271520 &  0.288630 &  0.281600 \\
    vqz    &  0.240890 &  0.250440 &  0.248190 &  0.248940 &  0.272290 &  0.289660 &  0.282990 \\
    v5z    &  0.240980 &  0.250700 &  0.248010 &  0.249050 &  0.272580 &  0.289970 &  0.283570 \\
    \noalign{\vskip 1mm}
    cbs23  &  0.241474 &  0.251180 &  0.249479 &  0.251269 &  0.273481 &  0.291015 &  0.286520 \\
    cbs34  &  0.240996 &  0.250583 &  0.247369 &  0.249413 &  0.272957 &  0.290555 &  0.283961 \\
    cbs345 &  0.241146 &  0.250876 &  0.247735 &  0.249337 &  0.273071 &  0.290558 &  0.284041 \\
    cbs45  &  0.241346 &  0.251265 &  0.248222 &  0.249235 &  0.273222 &  0.290562 &  0.284147 \\
    exper  &  0.24113  &  0.25093  &  0.24792  &  0.24866  &  0.27320  &  0.29041  &  0.28394  \\
    \end{tabular}
    \end{ruledtabular}
    
\end{table}

\begin{table}
\caption{SHCI dissociation energies for the molecules considered in this work.
    cbs23 indicates an extrapolation using only vdz and vtz bases, and cbs34/cbs45/cbs345 are labeled in the same way.}
    \label{tab:be_basis_extrap}
    \begin{ruledtabular}
\begin{tabular}{lrrrrrrr}
 & \multicolumn{7}{c}{Dissociation Energy (Ha)} \\
basis &       ScO &       TiO &        VO &       CrO &       MnO &       FeO &       CuO \\
\hline
vdz    &  0.241940 &  0.239450 &  0.226850 &  0.165090 &  0.127000 &  0.136620 &  0.104880 \\
vtz    &  0.250810 &  0.249200 &  0.239260 &  0.170730 &  0.138600 &  0.151240 &  0.106120 \\
vqz    &  0.255364 &  0.253398 &  0.243812 &  0.173056 &  0.142971 &  0.155512 &  0.108734 \\
v5z    &  0.256841 &  0.254443 &  0.244394 &  0.174247 &  0.143774 &  0.156712 &  0.109878 \\
\hline
cbs23  &  0.254384 &  0.253086 &  0.244229 &  0.173665 &  0.143113 &  0.156775 &  0.106230 \\
cbs34  &  0.258321 &  0.256149 &  0.246622 &  0.174208 &  0.145724 &  0.158243 &  0.110471 \\
cbs345 &  0.258330 &  0.255835 &  0.245885 &  0.174716 &  0.145210 &  0.158082 &  0.110693 \\
cbs45  &  0.258343 &  0.255415 &  0.244903 &  0.175394 &  0.144526 &  0.157868 &  0.110989 \\
\end{tabular}
\end{ruledtabular}

\end{table}

%\begin{table}
%    \caption{RMS deviations of the ionization energies and dissociation energies with respect to the extrapolation from vqz and v5z (cbs45) of the SHCI dissociation energies.
%    {\color{red} May be add column for total energy too.}
%    }
%    \label{tab:rms_ip_be_basis_extrap}
%    \begin{ruledtabular}
%    \begin{tabular}{lcc}
%          & \multicolumn{2}{c}{RMS Deviation (Ha)} \\
%    basis & Ionization energy & Dissociation energy  \\
%    \hline
%        vdz &  0.005572 &  0.015087 \\
%        vtz &  0.001442 &  0.005926 \\
%        vqz &  0.000657 &  0.002085 \\
%        v5z &  0.000448 &  0.001021 \\
%    \noalign{\vskip 1mm}
%      cbs23 &  0.000942 &  0.002280 \\
%      cbs34 &  0.000360 &  0.000822 \\
%     cbs345 &  0.000206 &  0.000469 \\
%\end{tabular}
%\end{ruledtabular}
%\end{table}

\begin{table}
\caption{RMS deviations of the SHCI total, ionization, and dissociation energies for various basis sets and extrapolations
with respect to the cbs45 extrapolation.
}
\label{tab:rms_ip_be_basis_extrap}
\begin{ruledtabular}
\begin{tabular}{lccc}
      & \multicolumn{3}{c}{RMS Deviation (Ha)} \\
basis & Total energy & Ionization energy & Dissociation energy \\
\hline
    vdz & 0.169949 &  0.007215 &  0.015820 \\
    vtz & 0.075498 &  0.001572 &  0.005997 \\
    vqz & 0.034565 &  0.000756 &  0.002160 \\
    v5z & 0.017665 &  0.000482 &  0.001063 \\
\noalign          {\vskip 1mm}
  cbs23 & 0.036296 &  0.001290 &  0.002683 \\
  cbs34 & 0.005110 &  0.000455 &  0.000981 \\
 cbs345 & 0.002919 &  0.000260 &  0.000561 \\
\end{tabular}
\end{ruledtabular}
\end{table}

\clearpage
\bibliography{realistic}